%% file: protctTree-J-Lin_Zhili-L.tex
\documentclass[conference,10pt,letterpaper]{IEEEtran}

\ifCLASSINFOpdf
\else
\fi

\usepackage{graphicx, bbm, latexsym, amsfonts, verbatim, amsmath, amssymb, algorithm, algorithmic, color, verbatim}
\usepackage{color, url}
\usepackage{tikz}
\usepackage{multirow}
\usepackage{array,booktabs}
\usepackage{epstopdf}
\usepackage{caption}
\usepackage{subcaption}

\newtheorem{corollary}{Corollary}
\newtheorem{define}{Definition}

\newtheorem{lemma}{Lemma}

\newtheorem{problem}{Problem}

\newtheorem{proposition}{Proposition}

\newtheorem{theorem}{Theorem}

\newcommand{\zred}{\textcolor{black}}

\def\BEA{\begin{eqnarray}}
\def\EEA{\end{eqnarray}}

\def\BEAN{\begin{eqnarray*}}
\def\EEAN{\end{eqnarray*}}

\def\BE{\begin{equation}}
\def\EE{\end{equation}}

\begin{document}
%
\title{Survivable Probability of Network Slicing with Random Physical Link Failure}

\author{\IEEEauthorblockN{Zhili~Zhou}
\IEEEauthorblockA{United Airlines\\
Chicago, IL USA\\
Email: zhilizhou@gmail.com}
\and
\IEEEauthorblockN{Tachun~Lin}
\IEEEauthorblockA{Department of Computer Science and Information Systems\\
Bradley University, Peoria, IL USA\\
Email: djlin@bradley.edu}
}
\maketitle

\begin{abstract}\label{abstract}
The fifth generation of communication technology (5G) revolutionizes mobile networks and the associated ecosystems through the integration of cross-domain networks. Network slicing is an enabling technology for 5G as it provides dynamic, on-demand, and reliable logical network slices (i.e., network services) over a common physical network/infrastructure.
Since a network slice is subject to failures originated from disruptions, namely node or link failures, in the physical infrastructure, our utmost interest is to evaluate the reliability of a network slice before assigning it to customers.
In this paper, we propose an evaluation metric, \textit{survivable probability}, to quantify the reliability of a network slice under random physical link failure(s). We prove the existence of a \textit{base protecting spanning tree set} which has the same survivable  probability as that of a network slice. We propose the necessary and sufficient conditions to identify a base protecting spanning tree set and develop corresponding mathematical formulations, which can be used to generate reliable network slices in the 5G environment. In addition to proving the viability of our approaches with simulation results, we also discuss how our problems and approaches are related to the Steiner tree problems and present their computational complexity and approximability.

\end{abstract}


\section{Introduction}

5G communications ``empower socio-economic transformation in countless ways, including those for productivity, sustainability, and well-being''~\cite{ngmn2016whitepaper5G}. The latest optical techniques~\cite{liu2016emerging}\cite{wang2016handover} and architectures~\cite{iovanna2016future}\cite{assimakopoulos2016switched}
serve as the global network infrastructure which provides capacities and guarantees the performance of 5G networks, especially network diversity, availability, and coverage.

To satisfy the requirements of different subscriber types, applications, and use cases, network slicing was introduced which enables programmability of network instances called \textit{network slices}. These instances should satisfy the bilateral service level agreement (SLA)~\cite{5gamerica2016network}\cite{ngmn2016netSlicingSecurity}, such as latency, reliability, and value-added services, among \zred{virtual} network operators and subscribers, \zred{especially mobile operators and subscribers}, in 5G systems.
\zred{Network slicing allows multiple virtual networks to be created on top of a common underlying physical infrastructure (including physical and/or virtual networks)~\cite{ngmn2016description}.}
Since the instantiation of network slices involves \zred{a physical network and multiple virtual networks}, a general way to model such networks is through the cross-layer network topologies.
The reliability of the physical infrastructure directly affects the network capabilities and performance level \zred{that} a network slice can provide.
Thus, a way to identify and quantify the reliability of a \zred{cross-layer network} when disruptions occur to the physical infrastructure, \zred{which leads to more reliable network slicing}, would be of interest to the virtual network operators.

To design a reliable cross-layer network, a key question to be answered is how to quantify the reliability of a cross-layer network.
When considering the reliability of single-layer networks, link failures are described as random events with corresponding failure probabilities, and the \textit{survivable probability} is the probability of a network to remain connected after random physical link failure(s)~\cite{yallouz2014tunable}\cite{yallouz2017tunable}.
Comparatively, a failure in the physical infrastructure of a cross-layer network may not only disrupt the flows in the physical network, but also affect demands satisfaction in the network slices as the demands from each slice are routed/realized through the physical infrastructure.
In this paper, we assume that each physical link may carry its own probability of failure (reliability index) and introduce the concept of \textit{cross-layer network survivable probability} to capture the probability of \zred{virtual networks of} a network slice to remain connected after any physical link failure. In the rest of the paper, we'll use \textit{survivable probability} as an abbreviation for cross-layer network survivable probability.

Different from prior research on the survivable cross-layer network design where all physical links have either 0\% or 100\% probability of failure, the survivable probability concept offers network operators a way to fine-tune a \zred{cross-layer network} with the corresponding level of SLA before offering it to the subscriber. This concept can also be applied to several related applications, such as the design of reliable cloud~\cite{yang2015software} and IP-over-WDM~\cite{develder2012optical} networks, \zred{where an IP-over-WDM network carries the traffic of each IP link through a lightpath in the WDM network, which utilizes a single wavelength through optical nodes like OXCs and OADMs without opto-electro-optical (O-E-O) conversion on intermediate optical nodes; and a cloud network constructed on top of a data center network is connected by fiber optics.}


\section{Literature Review}\label{sec:lr}
The design of a \zred{reliable} single-layer network has two main mechanisms, namely protection and restoration~\cite{cholda2005network}\cite{heegaard2009network}\cite{ramamurthy1999survivable} which guarantee the network's connectivity after the failure(s) of network component(s). Two lines of investigation were conducted in the fields of operations research and telecommunication networks.~\cite{grotschel1995design}\cite{dahl1998cutting}\cite{smith2008algorithms}\cite{botton2013benders} explored mixed-integer programming techniques and proposed solution approaches for the survivable network design problem \zred{(with 100\% survivable probability)} through polyhedron studies. \zred{They usually do not consider network failures as random events but with 0\% or 100\% \zred{reliable probability.}}~\cite{koster2003demand}\cite{shaikh2011anycast}\cite{dzida2008path}\cite{orlowski2012complexity} studied the reliable optical network and optical routing design through $p$-cycles, any-cast routing, and path set protection.~\cite{floyd1997reliable}\cite{li2013scaling}\cite{biswas2004opportunistic} discussed reliable wireless network design with scalable reliable multicast protocols and opportunistic routing in multi-hop wireless networks.~\cite{tarique2009survey}\cite{sara2014routing}\cite{liu2016survey}\cite{salayma2017wireless} reviewed works on reliable mobile networks emphasizing multipath or position-based routing in mobile ad hoc, wireless sensor, and vehicular ad hoc networks.

The studies of cross-layer networks \zred{focusing on their survivable design}, an $\mathcal{NP}$-complete problem~\cite{garey79}\cite{ModNar01}, consider both logical and physical networks, where logical nodes and links are mapped onto physical nodes and paths, respectively (with different routing schemes).~\cite{KurThi05}\cite{TodRam07}\cite{parandehgheibi2014survivable} utilized a sufficient condition, disjoint mappings of logical links, for survivable cross-layer network design, which transformed the cross-layer network design problem into the single-layer setting. 
Necessary and sufficient conditions for survivable cross-layer network design were proposed in
~\cite{ModNar01}\cite{lee2011cross}\cite{rahman2013svne} via cross-layer cutsets, which require the enumeration of all cross-layer cutsets.
To avoid the enumeration,~\cite{zhou2017survivable}\cite{zhou2017novel} proposed another necessary and sufficient conditions based on a cross-layer protecting spanning tree set (in short, protecting spanning tree set), which guarantee the connectivity of the logical network through the existence of a protecting spanning tree after any physical link failure.
It has been shown theoretically and computationally that a survivable cross-layer routing/network design may not exist for a given network; its existence highly relies on the network topology. Thus, unless some specific network structure which guarantees survivability is embedded in a given network~\cite{KTLin10}, the analysis and study on how to quantify and design a good/maximal partially survivable cross-layer routing also motivate this work. Survivable probability, an evaluation metric applicable to all cross-layer network topologies, is our attempt to address this problem in a general sense.


In this paper, we develop the \zred{survivable probability} of a cross-layer network, which describes the chance of a network slice to maintain its service against failure(s) in the physical infrastructure. Its single-layer counterpart, discussed in~\cite{yallouz2014tunable}\cite{yallouz2017tunable}, introduced the level of survivability which is ``a quantitative measure for specifying any desired level of survivability'' through survivable spanning trees. Our design and its single-layer counterpart share the same assumption that each physical link is associated with a probability of failure. Nevertheless, these two problems are fundamentally different due to their network settings.

Another related work in~\cite{lee2014maximizing} evaluated the reliability of a cross-layer network under random physical link failure by calculating the failure polynomials.
Our proposed approach differs from that in~\cite{lee2014maximizing} in three aspects: (1) we seek an exact solution approach with the objective to quantify the maximal survivable probability rather than an approximation through failure polynomials (which involves enumeration of cross-layer cutsets); (2) \zred{relieving from cross-layer cutset enumeration, our approach is scalable to larger size cross-layer networks;} 
and
(3) our approach can address both random or unified failure probabilities on physical links compared with the unified one in~\cite{lee2014maximizing}.

Our contributions in this paper are as follows.
(1) We define \zred{cross-layer network survivable probability}, an evaluation metric on the reliability of cross-layer networks.
(2) We demonstrate the existence of the protecting spanning tree set (as the \zred{base protecting spanning tree set}) which shares the same \zred{survivable probability} as that of a given cross-layer network. We prove the necessary and sufficient conditions to identify a base protecting spanning tree set.
(3) Our proposed approach, which requires at most $|E_P|$ (the number of physical links) protecting spanning trees, directly calibrates the survivable probability through a base protecting spanning tree set while avoiding the enumeration of cross-layer cutsets.
(4) By constructing a base protecting spanning tree set, the maximal survivable probability of a cross-layer network is tractable. Given a unified physical link failure probability, we prove that the design of a cross-layer network with the maximal survivable probability is equivalent to the cross-layer network design with the minimal number of shared physical links utilized by a base protecting spanning tree set.
(5) We prove that the maximal protecting spanning tree, a protecting spanning tree with the maximal \zred{survivable probability}, is a Steiner tree in the physical network whose terminal nodes are the corresponding physical nodes \zred{onto which the logical nodes are mapped}. We also discuss that the Steiner tree packing problem along with network augmentation may provide the maximal \zred{survivable probability} (100\%) in a cross-layer routing. 

The rest of this paper is organized as follows. Section~\ref{sec:problem} provides formal definitions and descriptions of the survivable probability and base protecting spanning tree set. Mathematical formulations for the maximal protecting spanning tree 
and the maximal survivable probability are presented in Section~\ref{sec:approach}. We discuss the relationship between the \zred{protecting spanning tree in a cross-layer network} and Steiner tree in a single-layer network in Section~\ref{sec:PST-Steiner}, followed by the simulation results in Section~\ref{sec:result} and conclusions in Section~\ref{sec:conclusion}.

\section{Definitions and Problem Description}\label{sec:problem}

Given a physical network denoted as $G_P=(V_P, E_P)$, and a logical network (i.e., \zred{a virtual network in a network slice}) denoted as $G_L=(V_L, E_L)$, where each logical node has an one-to-one mapping onto a physical node and each logical edge has an one-to-one mapping onto a physical path. \zred{We let $M(\cdot)$ denote the general logical-to-physical mapping function.} The logical-to-physical node mapping is denoted as $\zred{M}(s) = i$, $s\in V_L$ and $i\in V_P$; $\zred{M}(u)=p_u$, $u\in E_L$ and $p_u \subset E_P$ is the logical-edge-to-physical-path mapping; \zred{and $M(\tau)=\cup_{u\in \tau}M(u)$ is the mapping of a logical spanning tree $\tau\subset G_L$} onto $G_P$.
Notations and parameters used in this paper are listed in Table~\ref{tbl:notation}.
\begin{table}[t]
\begin{tabular}{p{2cm}|p{6cm}}
\hline\hline
 \rule{0pt}{9pt}Notation       &Description\\
\hline
 \rule{0pt}{8pt} $G_P = (V_P,E_P)$ &Physical network, where $V_P$ and $E_P$ represent the node and edge set, respectively, with node indices $i,j$ and link index $e$\\
 \rule{0pt}{8pt} $G_L = (V_L,E_L)$ &Logical network, where $V_L$ and $E_L$ denote the node and edge set, respectively, with node indices $s,t$ and link indices $\mu$,$\nu$\\
 \rule{0pt}{8pt} $(G_P,G_L)$ & The cross-layer network with known logical-to-physical mapping\\
 \rule{0pt}{8pt}$\mathcal{P}_{u}$ & A set of physical paths (routings) for $u\in E_L$, where $p_u$ is an element of $\mathcal{P}_{u}$, i.e., $p_u\in \mathcal{P}_u$\\
 \rule{0pt}{8pt} $T, \tau$& A protecting spanning tree set with $\tau$ as a protecting spanning tree, i.e., $\tau\in T$\\
 \rule{0pt}{8pt} $M(\cdot)$ & A general logical-to-physical mapping function, with node mapping $M(s) = i$, link mapping $M(\mu)=p_{\mu}$, and protecting spanning tree mapping $M(\tau)=\cup_{\mu\in \tau}p_{\mu}$\\
 \rule{0pt}{8pt} $\lambda$      &A tuple which denotes a protecting spanning tree and its mapping, i.e., $\lambda = [\tau, M(\tau)]$\\
 \rule{0pt}{8pt} $\Lambda$ & A tuple which denotes a protecting spanning tree set and its mapping, i.e., $\Lambda=\{\lambda\}$\\
 \rule{0pt}{8pt} $\Lambda^{F}(M(E_L))$ & A collection of protecting spanning tree's with link mapping $M(E_L)$\\
 \rule{0pt}{8pt} $\Lambda^{B}(G_P,G_L)$ & A base protecting spanning tree set and its mapping of a cross-layer network $G_P$ and $G_L$\\
 \rule{0pt}{8pt} $E_{P}(\lambda)$ &All physical links utilized by the routings of $\lambda$'s branches\\
 \rule{0pt}{8pt} $E^{M}_{P}(T)$ & Common physical links shared by the routings of all $\lambda\in T$\\
 \rule{0pt}{8pt} $\Omega(E_L)$       &A set of logical-to-physical link mappings, where $M(E_L)\in \Omega(E_L)$ is one of its instances\\
 \rule{0pt}{8pt} $R(M(E_L))$    &A set of physical links whose failures disconnect $G_L$ over a given mapping $M(E_L)$\\
 \rule{0pt}{8pt} $\Phi(G_P,G_L)$&The survivable probability of a cross-layer network\\
\hline
\hline
 \rule{0pt}{9pt}Parameter&Description\\
\hline
 \rule{0pt}{8pt} $\rho_e$       &Probability of failure for physical link $e$, $e\in E_P$\\
 \rule{0pt}{8pt} $\rho$         &Unified probability of failure for all $e\in E_P$\\
\hline\hline
\end{tabular}
\vspace{2pt}
\caption{Notations and parameters}
\label{tbl:notation}
\end{table}

\subsection{Protecting Spanning Tree Set}\label{subsec:protSpanningTree}
For a given logical-to-physical mapping $M(\cdot)$ of a cross-layer network $(G_P, G_L)$, the corresponding co-mapping~\cite{zhou2017survivable}, denoted as $M^{C}(\cdot)$, is defined as follows. Co-mapping of a logical edge $\nu$ is $M^{C}(\nu) = E_P \setminus M(\nu)$ with $\nu\in E_L$; and co-mapping of \zred{logical spanning tree} $\tau$ is $M^{C}(\tau) = E_P \setminus \underset{\nu\in \tau} \bigcup M(\nu)$; that is, $M^{C}(\tau) = \bigcap_{\nu\in \tau} M^{C}(\nu)$.

Given $M(\cdot)$, $M^C(\cdot)$, and a set of logical spanning trees $T$ of a cross-layer network $(G_P, G_L)$, the protecting spanning tree set~\cite{zhou2017survivable} is defined as follows.
If physical link $(i, j)$ is in $M^C(\tau)$, $\tau\in T$, then $\tau$ is called a \textit{protecting spanning tree} which \textit{protects} $(i, j)$. If for every physical link $(i, j)$, there exists a spanning tree in $T$ which protects $(i,j)$, then the routing is a \textit{survivable routing}, and $T$ is called a \textit{protecting spanning tree set} for survivable routing.
In this paper, given a protecting spanning tree $\tau$,
we let $\lambda=[\tau, M(\tau)]$ denote a protecting spanning tree and its  mapping, and $E_{P}(\lambda)=\{e: e\in \cup_{\mu\in \tau}p_{\mu}\}$ be the physical link set utilized by the routings of \zred{$\lambda$}. 


Given these definitions, we may now derive the evaluation metric, the survivable probability, in the following section.

\subsection{Survivable Probability}\label{subsec:definition}
Given a cross-layer network ($G_P, G_L)$ and its node mapping $M(\nu)$ for all $\nu\in V_L$. We assume that each physical link $e\in E_P$ is associated with probability of failure $\rho_e$, where $0\leq \rho_e\leq 1$. The survivable probability of $(G_P, G_L)$ is defined as follows.
\begin{define}\label{def:netSp}
Given $(G_P, G_L)$ and the failure probability $\rho_e$, for all $e\in E_P$, the survivable probability of this network is the probability of the logical network to remain connected after any physical link failure(s).
\end{define}
Given a logical link $\mu\in E_L$ and its mapping $M(\mu) = p_{\mu}$, the survivable probability of $\mu$ is $\text{Prob}(\mu)=\prod_{e\in p_{\mu}}(1-\rho_e)$. Similarly, the survivable probability of a logical spanning tree $\tau$ is defined below.
\begin{define}\label{def:proTreeSp}
Given a cross-layer network $(G_P, G_L)$, a protecting spanning tree and its mapping $\lambda=[\tau, M(\tau)]$, the survivable probability of $\lambda$ is $\text{Prob}(\lambda)=\prod_{e\in E_{P}(\lambda)}(1-\rho_e)$.
\end{define}
The \textit{maximal protecting spanning tree} \zred{is a protecting spanning tree with one of its possible mappings that provide the maximal survivable probability, which is greater than or equal to the survivable probability of any other trees and their mappings.}

We now demonstrate how a protecting spanning tree set can be used to improve the survivable probability even with a given logical-to-physical mapping.
Let $T=\{\tau\}$ be a protecting spanning tree set, $\Lambda=\{\lambda\}$ be the set of \zred{protecting spanning tree and its mappings}, and $E^{M}_{P}(\Lambda)=\cap_{\lambda_i\in \Lambda}E_{P}(\lambda_i)$ be the common physical links utilized by the routings of $\lambda_i\in \Lambda$.

\begin{figure}[!h]
\centering
\includegraphics[scale=0.45]{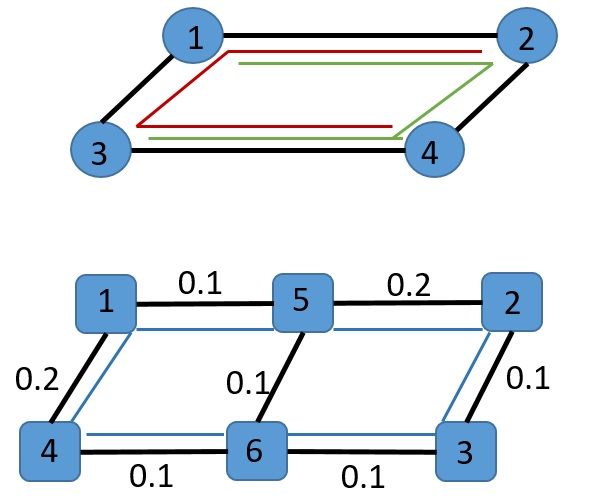}
\caption{survivable probability of a protecting spanning tree set}\label{fig:2tree}
\end{figure}

We use Fig.~\ref{fig:2tree} as an instance to illustrate the concept of a protecting spanning tree set and survivable probability. Given $G_L$ (top), $G_P$ (bottom), and $\rho_e, e\in E_P$ (labeled on each physical link). Logical-to-physical link mappings are given as follows:
$M(1,2)=\{(1,5),(5,2)\}$,
$M(1,3)=\{(1,4),(4,6),(6,3)\}$,
$M(2,4)=\{(2,3),(3,6),(6,4)\}$,
$M(3,4)=\{(3,6),(6,4)\}$.
\begin{table}[!h]
\begin{tabular}
{
>{\centering\arraybackslash}m{1.1cm}
>{\centering\arraybackslash}m{0.3cm}
>{\raggedright\arraybackslash}m{2.2cm}
>{\raggedright\arraybackslash}m{0.9cm}
>{\raggedright\arraybackslash}m{2.5cm}
}
\hline\hline
 \rule{0pt}{9pt} &$\tau$ & $M(\tau)$ &$E_P(\lambda)$ &Prob($\lambda$)\\
\hline
 \rule{0pt}{10pt} Red $\lambda_1\quad$ $[\tau_1,M(\tau_1)]$&(1,2), (1,3), (3,4)&$\{(1,5),(5,2)\}$; $\{(1,4),(4,6),(6,3)\}$; $\{(4,6),(6,3)\}$&\{(1,4),(1,5), (2,5),(3,6), (4,6)\} &$\prod_{e\in E_P(\lambda_1)}(1-\rho_e)$ = (1-0.2) (1-0.1) (1-0.2) (1-0.1) (1-0.1) = 0.46656\\
 \rule{0pt}{10pt}Green $\lambda_2\;$ $[\tau_2,M(\tau_2)]$&(1,2), (2,4), (4,3)&$\{(1,5),(5,2)\}$; $\{(2,3),(3,6),(6,4)\}$; $\{(4,6),(6,3)\}$&\{(1,5),(2,3), (2,5),(3,6), (4,6)\} &$\prod_{e\in E_P(\lambda_2)}(1-\rho_e)$ = (1-0.1) (1-0.2) (1-0.1) (1-0.1) (1-0.1)=\zred{0.52488}\\
\hline\hline
\end{tabular}
\vspace{2pt}
\caption{Protecting spanning trees and their mappings}
\label{tbl:instance}
\end{table}
We select a set of two protecting spanning trees: (red tree) $\lambda_1=[\tau_1, M(\tau_1)]$ and (green) $\lambda_2=[\tau_2, M(\tau_2)]$, whose branches, link mappings, utilized physical link sets, and survivable probability are presented in Table~\ref{tbl:instance}.
When considering a protecting spanning tree set and its mappings $\Lambda=\{\lambda_1, \lambda_2\}$, the common physical links used by the routings of both trees are $E^{M}_{P}(\Lambda)=\cap_{\lambda_i\in \Lambda}E_P(\lambda_i)=\{(1,5),(2,5),(3,6),(4,6)\}$. Therefore, any failure(s) occur among these links would disconnect both $\lambda_1$ and $\lambda_2$. Hence, the survivable probability of $\Lambda= (1-0.1)(1-0.2)(1-0.1)(1-0.1) = 0.5832$ which is higher than that of either $\lambda_1$ or $\lambda_2$. Derived from the example above
we have the following definition.
\begin{define}\label{def:treeSetSp}
Given a cross-layer network $(G_P, G_L)$, failure probability $\rho_e, e\in E_P$, a protecting spanning tree set and its mappings $\Lambda=\{\lambda\}$, the survivable probability of $\Lambda$ is $\text{Prob}(\Lambda)=\prod_{e\in E^{M}_{P}(\Lambda)}(1- \rho_{e})$.
\end{define}
We also define the \textit{maximal protecting spanning tree set}
as a protecting spanning tree set with the maximal survivable probability given any logical link mappings.
\subsection{Survivable Probability, Link Mapping, and Base Protecting Spanning Tree Set}\label{subsec:netTreeSetRlt}
Given a cross-layer network $(G_P, G_L)$, and mappings of all logical links $M(E_L)=\{M(\mu):  M(\mu)=p_{\mu}, p_{\mu}\in \mathcal{P}_{\mu}, \mu\in E_L\}$. Let $\Omega(E_L)=\{M(E_L)\}$ be the set of all logical link mappings, \zred{i.e.,} $\Omega(E_L)$ contains all possible combinations of logical link mappings for all logical links.
In this section, we explore the relation among $(G_P, G_L)$, $M(E_L)$, and protecting spanning tree set $T$. We demonstrate that the existence of the maximal protecting spanning tree set whose survivable probability is the same as \zred{the maximal survivable probability} of $(G_P, G_L)$. We also provide the necessary and sufficient conditions to identify such a $T$ and then evaluate the survivable probability accordingly.
We denote the maximal survivable probability of $(G_P, G_L)$ as $\Phi(G_P, G_L)$.
\begin{proposition}\label{prop:cldNetSurProb}
Given a cross-layer network $(G_P, G_L)$, all possible logical link mappings $\Omega(E_L)$, and failure probability $\rho_e$, $e\in E_P$. The maximal survivable probability of $(G_P, G_L)$, $\Phi(G_P, G_L) = \max_{M(E_L)\in \Omega(E_L)}\prod_{e\in R(M(E_L))}(1-\rho_e)$, where $R(M(E_L))$ denotes a set of physical links whose failure(s) disconnect $G_L$ with $M(E_L)$.
\end{proposition}
\begin{IEEEproof}
With Definition~\ref{def:netSp}, $(G_P, G_L)$'s survivable probability is determined by physical links whose failures disconnect $G_L$. For given logical link mappings $M(E_L)$, $R(M(E_L))$ contains all physical links whose failure(s) disconnect $G_L$. Hence, $G_L$ remains connected if and only if none of the links in $R(M(E_L))$ fail. Hence, $\prod_{e\in R(M(E_L))}(1-\rho_e)$ provides the survivable probability for $(G_P, G_L)$ over a given mapping $M(E_L)$. If $\Omega(E_L)=\{M(E_L)\}$ contains all possible combinations of the logical-to-physical link mappings, $\Phi(G_P, G_L)=\max_{M(E_L)\in \Omega(E_L)}\prod_{e\in R(M(E_L))}(1-\rho_e)$, which provides the maximal survivable probability of $(G_P, G_L)$.
\end{IEEEproof}
With Proposition~\ref{prop:cldNetSurProb}, $(G_P, G_L)$'s survivable probability is determined by its logical link mapping. We let $M^{*}(E_L)$ denote the logical link mapping which provides the maximal survivable probability for $(G_P, G_L)$, i.e., $M^{*}(E_L)=\arg_{M(E_L)\in \Omega(E_L)}\Phi(G_P, G_L)$.
\begin{theorem}\label{thm:existTreeSet}
Given a cross-layer network $(G_P, G_L)$ and failure probability $\rho_e$, $e\in E_P$, there exists a protecting spanning tree set and its mapping whose survivable probability is the same as that of $(G_P, G_L)$.
\end{theorem}
Please refer to Appendix~\ref{app:existTreeSet-proof} for the proof of Theorem~\ref{thm:existTreeSet}.
With Theorem~\ref{thm:existTreeSet}, we define the protecting spanning tree set which provides the maximal survivable probability, i.e., $\Phi(G_P, G_L)$, as a \textit{base protecting spanning tree set}.
We let $\Lambda^{B}(G_P,G_L)$ represent a base protecting spanning tree set and its mappings of a given cross-layer network $(G_P, G_L)$.

\begin{corollary}\label{cor:ncSufBaseProTreeSet}
A protecting spanning tree set is a base protecting spanning tree set if and only if it is the maximal protecting spanning tree set.
\end{corollary}
\begin{IEEEproof}
Proof of the necessary condition: follow the proof in Theorem~\ref{thm:existTreeSet}.\\
Proof of the sufficient condition: by contradiction.
If there exists a protecting spanning tree set with higher survivable probability than that of the base protecting spanning tree set, it leads to higher survivable probability. Contradiction!
\end{IEEEproof}
\zred{With Lemma~\ref{lm:flLink} in Appendix~\ref{app:existTreeSet-proof}, given a survivable routing, every physical link is protected by a least one protecting spanning tree. A base protecting spanning tree set also protects all physical links. In other word, $E^{M}_{P}(E_L)=\emptyset$.
Hence, a survivable cross-layer network has 100\% survivable probability against arbitrary physical link failure.}
Therefore, the survivable cross-layer network design problem with guaranteed 100\% survivable probability is a subproblem of the \zred{cross-layer network design with maximal survivable probability}.

\subsection{Unified Physical Link Failure Probability}\label{subsec:failProb}
Unified probability of failure, where the failure probability for all physical links is the same (i.e., $\rho_e=\rho$ with $e\in E_P$), is a special case of \zred{random physical link failure}.
We first study the maximal protecting spanning tree and have the following conclusion.
\begin{proposition}~\label{prop:mxTreeSpCase}
Given $(G_P, G_L)$ and physical link failure probability, $\rho_e=\rho$ with $e\in E_P$, the maximal protecting spanning tree is a tree $\tau$ with the minimal number of physical links utilized in $M(\tau)$.
\end{proposition}
\begin{IEEEproof}
Based on Definition~\ref{def:proTreeSp}, the maximal survivable probability \zred{of $\tau$ and its mapping is} $\max \prod_{e\in E_{P}(\lambda)}(1-\rho_e)$ $ = \max (1-\rho)^{|E_{P}(\lambda)|}$  (because $\rho_{e} = \rho$). Thus, $\lambda$ with $\min |E_{P}(\lambda)|$ will produce the maximal survivable probability.
\end{IEEEproof}
\begin{theorem}~\label{thm:failProbMinLk}
Given $(G_P, G_L)$ and unified failure probability $\rho$, a base protecting spanning tree set and its mapping with $\min|E^{M}_{P}(\Lambda^{B}(G_P,G_L))|$ \zred{provides the maximal} survivable probability of $(G_P, G_L)$ as $(1-\rho)^{\min|E^{M}_{P}(\Lambda^{B}(G_P,G_L))|}$.
\end{theorem}
\begin{IEEEproof}
With Definition~\ref{def:treeSetSp}, the survivable probability of a base protecting spanning tree set and its mapping is $\text{Prob}(\Lambda^{B}(G_P,G_L))=\prod_{e\in E^{M}_{P}(\Lambda^{B}(G_P,G_L))}(1-\rho_e)=(1-\rho)^{|E^{M}_{P}(\Lambda^{B}(G_P,G_L))|}$. According to Corollary~\ref{cor:ncSufBaseProTreeSet} and Proposition~\ref{prop:mxTreeSpCase}, 
minimizing $|E^{M}_{P}(\Lambda^{B}(G_P,G_L))|$ leads to $\Phi(G_P, G_L)$\zred{,} the maximum $\text{Prob}(\Lambda^{B}(G_P,G_L))$.
\end{IEEEproof}
Based on Theorem~\ref{thm:failProbMinLk}, finding survivable probability of $(G_P, G_L)$ with unified failure probability is equivalent to solving a cross-layer network design problem targeting the minimal number of shared physical links in its logical link mappings. The above proof also demonstrates that a base protecting spanning tree set and its mappings can provide a (partially) survivable network design along with a more precise evaluation metric on its reliability. Note that Theorem~\ref{thm:failProbMinLk} only holds when all physical links have the unified failure probability. If considering random link failure probabilities, the minimal set of physical links whose failures disconnect the logical network may not be equivalent to $E^{M}_{P}(M(E_L))$, thus leads to a survivable probability different from that of the cross-layer network.

Compared with the approach considered in~\cite{lee2014maximizing} where the reliability of a cross-layer network is approximated through the failure polynomials generated by enumerating exponential number of cross-layer cutsets  with unified link failure probability, the base protecting spanning tree set can provide \zred{exact calibration of survivable probability} under both unified and random physical link failure probabilities. 

\section{Solution Approach}\label{sec:approach}
Based on Theorems~\ref{thm:existTreeSet} and~\ref{thm:failProbMinLk}, we present in this section the mathematical programming formulations to compute survivable probability of a cross-layer network. 
We first present the formulation for survivable probability with unified physical link failure probability as a special case in Section~\ref{subsec:speCase}, followed by a generalized formulation addressing random probabilities of failure in Section~\ref{subsec:baseTreeSet}.
Variables and parameters used in the formulations are listed in Table~\ref{tbl:vrPr}.
\begin{table}[!b]
\begin{tabular}{p{2cm}|p{6cm}}
\hline\hline
 \rule{0pt}{9pt} Variable &Description\\
\hline
\rule{0pt}{8pt} $x_{ij}$& Binary variable indicating whether $(i,j)$'s failure disconnects the logical network. If yes, $x_{ij}=1$; otherwise, $x_{ij}=0$\\
 \rule{0pt}{8pt} $y^{st}_{ij}$& Binary variable indicating whether logical link $(s,t)$ is routed through physical link $(i,j)$ or not. If yes, $y^{st}_{ij}=1$; otherwise, $y^{st}_{ij}=0$\\
 \rule{0pt}{8pt} $z_{st}$& Binary variable indicating whether logical link $(s,t)$ is connected and forms a protecting spanning tree. If yes, $z_{st}=1$; otherwise, $z_{st}=0$\\
 \rule{0pt}{8pt} $w^{ij}_{st}$& Binary variable indicating whether logical link $(s,t)$ is connected and forms a protecting spanning tree after physical link $(i,j)$ failed. If yes, $z^{ij}_{st}=1$; otherwise, $z^{ij}_{st}=0$\\
 \rule{0pt}{8pt} $g_{ij}$& Binary variable indicating whether physical link $(i,j)$ is shared by trees in a base protecting spanning tree set. If yes, $g_{ij}=1$; otherwise, $g_{ij}=0$\\ \hline\hline
 \rule{0pt}{9pt} Parameter & Description\\
\hline
 \rule{0pt}{8pt} $c_{ij}$  & The coefficient for physical link. With unified failure probability, $c_{ij}=1$; with random physical link failure probability, $c_{ij}=\zred{-}\ln(1-\rho_{ij})$\\
\hline\hline
\end{tabular}
\vspace{2pt}
\caption{Variables and parameters used in mathematical formulations}
\label{tbl:vrPr}
\end{table}

\subsection{Survivable Probability of Cross-layer Networks with Unified Physical Link Failure Probability}\label{subsec:speCase}
We first present a mixed-integer programming formulation to generate the maximal protecting spanning tree, followed by a formulation to generate a base protecting spanning tree set.

\subsubsection{Maximal Protecting Spanning Tree}\label{subsubsec:maxTree}
Given unified failure probability $\rho$ on physical links, we propose a mixed integer programming formulation with the objective to minimize the number of physical links utilized in tree branches' routings (based on Proposition~\ref{prop:mxTreeSpCase}).
\begin{align}
& \min_{x,y,z}\sum_{(i,j)\in E_P}x_{ij}\nonumber\\
s.t.    &\sum_{(i,j)\in E_P}y^{st}_{ij}-\sum_{(j,i)\in E_P}y^{st}_{ji}=
\left\{\begin{matrix}
z_{st}, &\,\mbox{if } i=s,\\
-z_{st}, &\,\mbox{if } i=t,\\
0, &\,\mbox{if } i\neq \{s, t\},
&\end{matrix}\right.\label{fm:lightpath1}\\
        &y^{st}_{ij}+y^{st}_{ji}  \leq x_{ij}, (s,t)\in E_L, (i,j)\in E_P\label{fm:minLinkSel}\\
        &\sum_{(s,t)\in E_L}z_{st} - \sum_{(t,s)\in E_L} z_{ts} =
\left\{\begin{matrix}
|V_L|-1, \text{if } s=s_0\\
-1,      \text{if } s\neq s_0, s\in V_L
&\end{matrix}\right.\label{fm:maxTreeTree1}\\
        &\sum_{(s,t) \in E_L}z_{st} =|V_L|-1, (i,j)\in E_P \label{fm:zBnForced}\\
        &x_{ij}, y^{st}_{ij}, z_{st} \in \{0,1\}, (s,t)\in E_L, (i,j)\in E_P\label{fm:maxTreeRegion}
\end{align}
Constraint (\ref{fm:lightpath1}) maps logical links onto physical paths and \zred{selects logical links} forming a logical spanning tree, in which $z_{st}$ on the right hand side indicates whether $(s,t)$ is a branch of a logical spanning tree or not.
Constraint (\ref{fm:minLinkSel}) indicates which physical links are utilized by the routings of a selected protecting spanning tree.
Constraints (\ref{fm:maxTreeTree1}) and (\ref{fm:zBnForced}) form a protecting spanning tree corresponding to the logical link mapping generated by constraint (\ref{fm:lightpath1}). Constraint~(\ref{fm:maxTreeRegion}) provides the feasible regions for all decision variables.

\subsubsection{Base Protecting Spanning Tree Set}\label{subsubsec:treeSet}
The above formulation generates a maximal protecting spanning tree. Extending the formulation, we now present a mixed-integer programming formulation to compute the survivable probability with unified physical link failure probability.
Based on Corollary~\ref{cor:ncSufBaseProTreeSet} and Theorem~\ref{thm:failProbMinLk}, the proposed formulation has the objective to minimize the total number of physical links shared by $M(T)$ of a protecting spanning tree set $T$.
\begin{align}
\min_{\zred{g,w,y}} &\sum_{(i,j)\in E_P} g_{ij} \nonumber\\\
s.t. &\sum_{(i,j)\in E_P}y^{st}_{ij}-\sum_{(j,i)\in E_P}y^{st}_{ji}=
\left\{\begin{matrix}
\zred{1}, &\,\mbox{if } i=s,\\
\zred{-1}, &\,\mbox{if } i=t,\\
0, &\,\mbox{if } i\neq \{s, t\},
&\end{matrix}\right.\label{fm:lightpath2}\\
       &w^{ij}_{st}\leq 1-(y^{st}_{ij}+y^{st}_{ji}), (s,t)\in E_L, (i,j)\in E_P \label{fm:protTreeSet2}\\
       \sum_{(s,t)\in E_L}&w^{ij}_{st} - \sum_{(t,s)\in E_L} w^{ij}_{ts} =
\left\{\begin{matrix}
\zred{(1 - g_{ij})}, \qquad\text{if } s=s_0\\
\zred{(g_{ij}-1)/(|V_L|-1)},\;\;  \\
\qquad\quad\text{if } s\neq s_0, s\in V_L
&\end{matrix}\right.\label{fm:maxTreeTreeSet2}\\
       y^{st}_{ij}, &g_{ij} \in \{0,1\}, w^{ij}_{st}\geq 0, (s,t)\in E_L, (i,j)\in E_P\label{fm:maxTreeRegion2}
\end{align}
Similar to constraint (\ref{fm:lightpath1}), constraint (\ref{fm:lightpath2}) generates physical paths for logical links which are branches of a spanning tree in a base protecting spanning tree set. \zred{Constraints (\ref{fm:protTreeSet2})--(\ref{fm:maxTreeTreeSet2}) generate a protecting spanning tree after any physical link's failure if the physical link is protected; otherwise, the physical link is identified as unprotected. With the information of unprotected physical links, the generated protecting spanning tree set where each of its element protects at least one physical link is then identified as a base protecting spanning tree set.} Constraint~(\ref{fm:maxTreeRegion2}) provides the feasible regions for all decision variables.

\subsection{Survivable Probability of Cross-Layer Networks with Random Physical Link Failure Probability}\label{subsec:baseTreeSet}
In this section, we discuss a more generalized and realistic physical link failure scenario, where the physical link failure probability is not unique.
Based on Corollary~\ref{cor:ncSufBaseProTreeSet}, the objective used to select a base protecting spanning tree set is as follows.
\begin{align}
     &\max_{\Lambda}\prod_{_{e\in E^{M}_{P}(\Lambda)}}(1-\rho_e)\label{obj:nonlinear}
\end{align}
Constraint (\ref{obj:nonlinear}) is nonlinear. Applying the $\ln$ function to this constraint converts it into the linear form.
\begin{align}
\max_{\Lambda}\sum_{_{e\in E^{M}_{P}(\Lambda)}}\ln(1-\rho_e).\label{fm:objMaxWeight}
\end{align}
\zred{We let $c_{ij}$ be the weights on physical links. $c_{ij}=1$ for unified physical link failure probability, and  $c_{ij}=-\ln(1-p_{ij})$ for random probabilities of link failure.} The generalized formulation for the survivable probability of a cross-layer network is then
\begin{align}
\min_{y,g}&\sum_{(i,j)\in E_P}c_{ij}g_{ij}\nonumber\\
s.t. & \text{ Constraints (\ref{fm:lightpath2}) -- (\ref{fm:maxTreeRegion2})}
\end{align}
Note that \zred{with unified failure probability, the formulation for base protecting spanning tree set is to minimize} the total number of shared physical links. But with random failure probability, after linearization, the objective is to \zred{maximize the total weight of the shared physical links~\label{fm:objMaxWeight} with non-positive physical link weights}, because physical link failure probability is in $[0,1]$ and the value $\ln(1-p_e)$ \zred{is non-positive}. When we let the physical link weight $c_{ij}$ be $-\ln(1-p_{ij})$ with $(i,j)\in E_P$ \zred{as the non-negative weight, the generalized objective becomes minimizing the total weight (non-negative) of the shared physical links.}

\section{Protecting Spanning Tree v.s. Steiner Tree}~\label{sec:PST-Steiner}
In this section, we discuss the relationship between a protecting spanning tree in a cross-layer network and a Steiner tree in the physical network whose terminal nodes are the physical nodes corresponding to the logical nodes and Steiner nodes are a subset of the remaining physical nodes. First, we show that the maximal protecting spanning tree in a cross-layer network is a minimum Steiner tree in its physical network in which the terminal nodes are a subset of physical nodes onto which logical nodes are mapped. This conclusion leads to a factor $\ln 4 +\varepsilon$ approximation algorithm for the maximal protecting spanning tree in a cross-layer network. Motivated by the conclusion, we further study the relationship between survivable cross-layer network design and edge-disjoint Steiner tree packing problem (with 100\% survivable probability). We demonstrate that the existence of edge-disjoint Steiner tree packing with logical network augmentation provides necessary and sufficient conditions for survivable cross-layer routing. 
\subsection{Maximal Protecting Spanning Tree v.s. Minimum Steiner Tree}
The minimum Steiner tree problem~\cite{garey2002computers} is defined as follows.
\begin{problem}~\label{pb:mStp} The minimum Steiner tree problem~\cite{hauptmann2013compendium}\\
INSTANCE: Graph $G=(V,E)$, edge cost $c: E \rightarrow R^{+}$, a set of terminal nodes $S\subseteq V$. \\
SOLUTION: A tree $\gamma=(V_{\gamma}, E_{\gamma})$ in $G$ such that $E_{\gamma} \subseteq E$ and $S\subseteq V_{\gamma} \subseteq V$ 
with $s\in S$ \\
OBJECTIVE: Minimize cost function $\sum_{e\in E_{\gamma}}c(e)$.
\end{problem}
The minimum Steiner tree problem is $\mathcal{NP}$-hard~\cite{garey2002computers} and has a factor $(\ln 4 + \varepsilon)$ approximation algorithm~\cite{byrka2010improved}. Its special case in a planar graph is polynomial solvable in $\mathcal{O}(3^{k}n + 2^k (n\log n +m))$, where $n =|V|$, $k=|S|$, and $m=|E|$~\cite{borradaile2009n}.

We now demonstrate that the maximal protecting spanning tree problem is the minimum Steiner tree problem in a physical network. Let $V^{L}_{P}$ be a set of physical nodes which logical nodes are mapped onto.
\begin{theorem}\label{thm:maxPTminST}
Given a cross-layer network $(G_P, G_L)$, the maximal protecting spanning tree and its mapping $\lambda^{*}=[\tau^{*}, M(\tau^{*})]$. $M(\tau^{*})$ is the minimum Steiner tree in $G_P$ with $V^{L}_{P}$ as the terminal nodes and $c_e=-\ln(1-\rho_e)$ as the link costs.
\end{theorem}
The proof of Theorem~\ref{thm:maxPTminST} is given in Appendix~\ref{app:maxPTminST}.
\begin{corollary}\label{cor:mPTApp}
Given a cross-layer network $(G_P, G_L)$ and failure probability $\rho_e$ with $e\in E_P$, a factor $(\ln 4 + \varepsilon)$ approximation algorithm exists for the maximal protecting spanning tree problem.
If $G_P$ is a planner graph, a polynomial algorithm exists for the maximal protecting spanning tree problem.
\end{corollary}
Let $G_P=(V_P,E_P)$ be the physical network, $V^{L}_{P}$ be the terminal node set, and $V_P\setminus V^{L}_{P}$ be the superset of the Steiner node set. Each physical link is assigned a \zred{non-negative cost $c_e=-\ln(1-\rho_e)$} with $e\in E_P$.
Based on Theorem~\ref{thm:maxPTminST}, we can apply the factor $(\ln 4 + \varepsilon)$ approximation algorithm in~\cite{byrka2010improved}, and the maximal protecting spanning tree can be approximated by a $\ln 4 + \varepsilon$ factor.
Furthermore, a polynomial-time algorithm with complexity $\mathcal{O}(3^k n + 2^k (n \log n + m)$, where $n =|V_P|$, $k=|V^{L}_{P}|$, and $m=|E_P|$~\cite{borradaile2009n} exists for the maximal cross-layer protecting spanning tree problem, which only requires the physical network to be planar.

\subsection{Survivable Cross-layer Network Design with Augmentation v.s. Steiner Tree Packing}\label{subsubsec:maxSPCN}
Motivated by the construction of the maximal protecting spanning tree via a minimum Steiner tree in the physical network,
if considering \zred{multiple protecting spanning trees}, it leads us to the problem of packing edge-disjoint Steiner trees described below.
\begin{problem} Packing edge-disjoint Steiner trees~\cite{hauptmann2013compendium}
INSTANCE: An undirected multigraph $G=(V,E)$, and a set of terminal nodes $S\subseteq V$.\\
SOLUTION: A set $\Gamma=\{\gamma_1,\cdots, \gamma_m\}$ of Steiner trees $\gamma_i$ for $S$ in $G$ which have pairwise disjoint sets of edges.\\
OBJECTIVE: Maximize $|\Gamma|$.\\
\end{problem}
\cite{kaski2004packing} proved that finding two edge-disjoint Steiner trees is $\mathcal{NP}$-hard.
Next, we build the connection between survivable cross-layer network design with logical augmentation and Steiner tree packing. 
We define the link augmentation as follows.
\begin{define} \textit{Logical link augmentation}~\cite{zhou2017novel}\label{def:logAug}\\
Given a cross-layer network $(G_P, G_L)$ and a logical link $\mu=(s,t)\in E_L$. An augmented logical link $\mu'$ is a link parallel to $\mu$, and $M(\mu)$ and $M(\mu')$ are edge-disjoint.
\end{define}

\begin{theorem}\label{thm:logTreePhyTree}
Given a cross-layer network \zred{$(G_P, G_L)$}. Let $V^{L}_{P}$ be the set of physical nodes corresponding to the logical nodes. If 2 edge-disjoint Steiner trees are packed in $G_P$, where $V^{L}_{P}$ are the terminal nodes and $V_P\setminus V^{L}_{P}$ is the superset of the Steiner nodes, the survivability of the cross-layer routing is guaranteed with logical link augmentation.
\end{theorem}
\begin{IEEEproof}
Given a logical link $\mu=(s,t)\in G_L$, let $\mu'$ be the augmented logical link of $\mu$. With Definition~\ref{def:logAug}, $M(\mu)$ and $M(\mu')$ are edge-disjoint. 2 edge-disjoint Steiner trees in $G_P$ with $V^{L}_{P}$ as their terminal nodes guarantee the existence of two edge-disjoint paths $p_1$ and $p_2$ connecting $i$ and $j$ with $M(s)=i$, $M(t)=j$, and $i,j \in V^{L}_{P}$. Hence, after any physical link failure, $s$ and $t$ remain connected. Thus, the two edge-disjoint Steiner trees actually provide two protecting spanning trees in the logical network, which guarantee the connectivity of logical network after the failure of any physical link.
\end{IEEEproof}
With Theorem~\ref{thm:logTreePhyTree}, we have the following conclusions for the necessary condition to identify the survivability of a cross-layer network with logical link augmentation.
\begin{corollary}\label{col:augSurvNecCondition}
Given $G_P=(V_P,E_P)$, if $V_P\setminus V^{L}_{P}$ is 13-edge connected, then, 2 edge-disjoint Steiner trees exists.
\end{corollary}
The conclusion directly follows~\cite{west2012packing} that if the terminal nodes are 6.5$k$-edge connected, there exists $k$ edge-disjoint Steiner trees. Note here that two special cases require less edge connectivity on terminal nodes, namely $k$-regular graph~\cite{kriesell2009edge} and planar graph~\cite{aazami2012approximation}.

Furthermore, solution approaches solving edge-disjoint Steiner tree packing lead to solution approaches for survivable cross-layer routing design with logical link augmentation, which has a factor $\mathcal{O}(\sqrt{|V_P|}\log |V_P|)$ approximation algorithm~\cite{cheriyan2006hardness}.

Theorem~\ref{thm:logTreePhyTree} demonstrates that the cross-layer network design problem can be solved as its single-layer network counterpart with logical link augmentation. 
However, the same claim does not hold if the logical augmentation is not allowed.


\input{newSimResults-L}

\section{Conclusion}\label{sec:conclusion}
In this paper, we introduced a new evaluation metric, the survivable probability, to evaluate the probability of the logical network to remain connected against physical link failure(s) with either unified or random failure probabilities. We explored the exact solution approaches in the form of mathematical programming formulations. We also discussed the relationship between the survivable probability of a cross-layer network and the protecting spanning tree set, which led to the base protecting spanning tree set approach. We proved the existence of a base protecting spanning tree set in a given cross-layer network and its necessary and sufficient conditions. We demonstrated that cross-layer network survivability may be solved or approximated through the single-layer network structures with some techniques such as logical augmentation and some criteria such as planar graphs. Our simulation results showed the effectiveness of proposed solution approaches.

\begin{appendices}

\section{Proof of Theorem~\ref{thm:existTreeSet}}\label{app:existTreeSet-proof}
Given a cross-layer network $(G_P, G_L)$, a set of all logical-to-physical link mappings $\Omega(E_L)$, and \zred{a logical link mapping $M(E_L)\in \Omega(E_L)$. We let $\Lambda^{F}(M(E_L))=[\mathcal{T}^{F}, M(E_L)]$ be a protecting spanning tree set (containing all protecting spanning trees $mathcal{T}^{F}$) with logical link mapping $M(E_L)$.}
\begin{lemma}\label{lm:flLink}
Given $(G_P, G_L)$, $T^F$, and $\Lambda^{F}(M(E_L))$. $G_L$ remains connected after any physical link failure if and only if a protecting spanning tree $\tau$ exists which protects physical link $e$, with $e\in E_P, \tau\in T^{F}(M(E_L))$.
\end{lemma}
\begin{IEEEproof}
Proof of the necessary condition: given $M(E_L)\in\Omega(E_L)$, if $G_L$ remains connected after the failure of $e$, then, a logical spanning tree $\tau$ exists with branch mapping $M(\tau)\subset M(E_L)$. \\
Proof of sufficient condition: if a protecting spanning tree $\tau\in T^{F}$ protects $e$, then, $e\notin E_{P}(\tau)$. Hence, after $e$'s failure, $\tau$ guarantees the connectivity of $G_L$.  
\end{IEEEproof}
With Lemma~\ref{lm:flLink}, if $G_L$ is disconnected due to the failure of $e$, then, no protecting spanning tree exists to protect $e$ for the given $T^F$ and its mappings.
\begin{lemma}~\label{lm:eqSet}
For a logical link mapping $M(E_L)\in \Omega(E_L)$, $R(M(E_L))= E^{M}_{P}(\Lambda^{F}(M(E_L)))$.
\end{lemma}
\begin{IEEEproof}
We first prove that $R(M(E_L))\subseteq E^{M}_{P}(\Lambda^{F}(M(E_L)))$. Given $e\in R(M(E_L))$, with Lemma~\ref{lm:flLink}, no protecting spanning tree exists for $e$. Then, we have $e\notin E_P\setminus E_P(\lambda)$ with $\lambda\in \Lambda^{F}(M(E_L))$. Hence, $e\notin \cup_{\lambda\in \Lambda^{F}(M(E_L))}E_P\setminus E_P(\lambda)$. Let $A^{c}$ be the complement of set $A$. Then, $e\in [\cup_{\lambda\in \Lambda^{F}(M(E_L))}E_P\setminus E_P(\lambda)]^{c}$. We have $e\in \cap_{\lambda\in \Lambda^{F}(M(E_L))}E_P(\lambda)$. Therefore, $R(M(E_L))\subseteq E^{M}_{P}(\Lambda^{F}(M(E_L)))$.\\
We now prove that $E^{M}_{P}(\Lambda^{F}(M(E_L)))\subseteq R(M(E_L))$. Given a physical link $e\in E^{M}_{P}(\Lambda^{F}(M(E_L)))$, then, $e\in E_P(\lambda)$ for all $\lambda\in \Lambda^{F}(M(E_L))$. With Lemma~\ref{lm:flLink}, no protecting spanning tree protects $e$, hence, $e\in R(M(E_L))$. Therefore, $E^{M}_{P}(\Lambda^{F}(M(E_L)))\subseteq R(M(E_L))$.
\end{IEEEproof}
\textit{Theorem~\ref{thm:existTreeSet}:}
For a cross-layer network $(G_P,G_L)$, there exists a protecting spanning tree set which has the same survivable probability as that of $(G_P,G_L)$.
\begin{IEEEproof}
We let $M^{*}(E_L)$ be the logical link mapping with the maximal survivable probability, i.e., $M^{*}(E_L)=\arg_{M(E_L)\in \Omega(E_L)}\max\prod_{e\in R(M(E_L))}(1-\rho_e)$. Let
$M'(E_L)$ be the logical link mapping for the maximal survivable probability of a cross-layer spanning tree set, i.e., $M'(E_L)=\arg_{M(E_L)\in \Omega(E_L)}\max\prod_{e\in E^{M}_{P}(\Lambda(M(E_L)))}(1-\rho_e)$. We now prove that $M^{*}(E_L)=M'(E_L)$. \\
With Lemma~\ref{lm:eqSet}, we have $R(M^{*}(E_L))= E^{M}_{P}(\Lambda^{F}(M^{*}(E_L)))$ and $R(M'(E_L))= E^{M}_{P}(\Lambda^{F}(M'(E_L)))$.

With the definition of $M^{*}(E_L)$ and $M'(E_L)$, we have
$\prod_{e\in E^{M}_{P}(\Lambda^{F}(M'(E_L)))}(1-\rho_{e}) =\prod_{e\in R(M'(E_L))}(1-\rho_{e})\leq \prod_{e\in R(M^{*}(E_L))}(1-\rho_{e})$;
and $\prod_{e\in R(M^{*}(E_L))}(1-\rho_{e})=\prod_{e\in E^{M}_{P}(\Lambda^{F}(M^{*}(E_L)))}(1-\rho_{e}) \leq \prod_{e\in E^{M}_{P}(\Lambda^{F}(M'(E_L)))}(1-\rho_{e})$. Hence, $\prod_{e\in R(M^{*}(E_L))}(1-\rho_{e})=\prod_{e\in E^{M}_{P}(\Lambda^{F}(M^{*}(E_L)))}(1-\rho_{e})$. The conclusion holds.
\end{IEEEproof}

\section{Proof of Theorem~\ref{thm:maxPTminST}}\label{app:maxPTminST}
\begin{lemma}\label{lm:noCycleInMapping}
Given a cross-layer network $(G_P, G_L)$, and the maximal $\mathcal{PST}\ \tau^{*}$ and its mapping $\lambda^{*}=[\tau^{*}, M(\tau^{*})]$. $M(\tau^{*})=(V^{L}_{P}, E(M(\tau^{*})))$ is a tree in $G_P$.
\end{lemma}
\begin{IEEEproof}
We prove this conclusion by contradiction.
Given a maximal protecting spanning tree and its mapping, $\lambda^{*}$.
Since $\tau^{*}$ is a logical spanning tree and $M(\tau^{*})$ is its mapping onto $G_P$, physical nodes in $V^{L}_{P}$ are connected. With the maximal protecting spanning tree, we have $\tau^{*}=\arg_{\tau\in \{\tau\}}\zred{\max}\prod_{e\in E(M(\tau))}(1-\rho_{e})$. The $\zred{\max}\prod_{e\in E(M(\tau))}(1-\rho_{e})$ leads to $\min \sum_{e\in E(M(\tau))} c_e$. As discussed earlier, we consider $c_e=\zred{-}\ln(1-\rho_e)$ as the edge cost.

If $M(\tau^{*})$ is not a tree in $G_P$, then, at least a cycle exists in $M(\tau^{*})$, denoted as $\varsigma=(\zred{V^{L}_{P}}, E_\varsigma)\subseteq G_P$. By removing an edge subset of $\varsigma$, a spanning tree could be constructed with $\zred{V^{L}_{P}}$ remaining connected; otherwise, $\zred{V^{L}_{P}}$ is not fully connected in $M(\tau^{*})$, which contradicts the condition that $M(\tau^{*})$ is connected and with minimal weight (after removing edges in $\varsigma$). Hence, the conclusion holds.
\end{IEEEproof}
\begin{IEEEproof} [\textit{Proof of Theorem~\ref{thm:maxPTminST}}]
With Lemma~\ref{lm:noCycleInMapping}, $M(\tau^{*})$ connects $V^{L}_{P}$ without cycles. Taking $V^{L}_{P}$ as terminal nodes and $V_P\setminus V^{L}_{P}$ as the superset of Steiner nodes, $M(\tau^{*})$ constructs a spanning tree connecting all $V^{L}_{P}$ via nodes in $V_P \setminus V^{L}_{P}$ and edges in $E_P$. Meanwhile, with edge cost $c_e=\zred{-}\ln(1-\rho_e)$, $\sum_{e\in E(M(\tau^*))}c_{e}$ is  minimal as $\lambda^{*}=[\tau^{*}, M(\tau^*)]$ is the maximal protecting spanning tree and its mapping. Hence, the conclusion holds.
\end{IEEEproof}

\section{MIP Formulation for Survivable Cross-layer Network Routing}\label{app:MIP}
We utilize the following MIP formulation~\cite{zhou2017novel} (SUR-TEST) to test whether a a cross-layer network is survivable or not. \zred{The definitions of variables are in Table~\ref{tbl:vrPr}}. After executing the formulation, if a feasible solution exists, the cross-layer network is survivable; otherwise, the cross-layer network is non-survivable.
\begin{align}
\min &\sum_{(i,j)\in E_P} y_{ij}\nonumber\\
s.t.&\sum_{(i,j)\in E_P}y^{st}_{ij}-\sum_{(j,i)\in E_P}y^{st}_{ji}=
\left\{\begin{matrix}
\zred{1}, &\,\mbox{if } i=s,\\
\zred{-1}, &\,\mbox{if } i=t,\\
0, &\,\mbox{if } i\neq \{s, t\},
&\end{matrix}\right.\label{fm:lightpath2_Suv}\\
       &w^{ij}_{st}\leq 1-(y^{st}_{ij}+y^{st}_{ji}), (s,t)\in E_L, (i,j)\in E_P \label{fm:protTreeSet2_Suv}\\
       \sum_{(s,t)\in E_L}&w^{ij}_{st} - \sum_{(t,s)\in E_L} w^{ij}_{ts} =
\left\{\begin{matrix}
\zred{1}, \qquad\text{if } s=s_0\\
\zred{-1/(|V_L|-1)},\;\;  \\
\qquad\quad\text{if } s\neq s_0, s\in V_L
&\end{matrix}\right.\label{fm:maxTreeTreeSet2_Suv}\\
y^{st}_{ij}&\in \{0,1\}, w^{ij}_{st} \in [0,1], (s,t)\in E_L, (i,j)\in E_P
\end{align}

\end{appendices}

\bibliographystyle{IEEEtran}
\bibliography{IEEEabrv,pt2}

\end{document}

%% file: newSimResults-L.tex
\section{Simulation Study}\label{sec:result}
\zred{In this section, we present our simulation design, testing cases setup, simulation results and observations.}
The goal is to validate and demonstrate the effectiveness of the proposed base protecting spanning tree set in \zred{calibrating} the survivable probability which supporting network slicing over small and median-size cross-layer networks.
\subsection{Objectives for Simulations}\label{subsec:simDesign}
\zred{The testing cases and simulations are designed to verify that (1) given a survivable cross-layer network, our base protecting spanning tree set approach should provide 100\% survivable probability regardless of the probability of failure on physical links; (2) with unified failure probability, the minimal number of shared physical links in the logical-edge-to-physical-path mappings result in the same survivable probability as that of the base protection spanning tree set; (3) the maximal protecting spanning tree provides a lower bound estimation for the survivable probability of a cross-layer network; 
also, we want to know how tight the lower bound estimation performs numerically; and (4) the survivable probability can be an evaluation metric for both survivable and non-survivable networks 
with either unified or random probabilities of failure on physical links. Last but not least, we want to observe and report the behaviors between survivable and non-survivable cross-layer networks with either uniform or random failure probabilities, which may provide insights/directions for future studies.}

\subsection{Simulation Setup}\label{subsec:testCase}
Based on the objectives above, we now present the selection of small and medium size cross-layer networks, failure probabilities, and the composition of testing cases.
\subsubsection{Small Size Cross-layer Network with NSF as the Physical Network}
\begin{figure}[!t]
\centering
\includegraphics[scale=0.43]{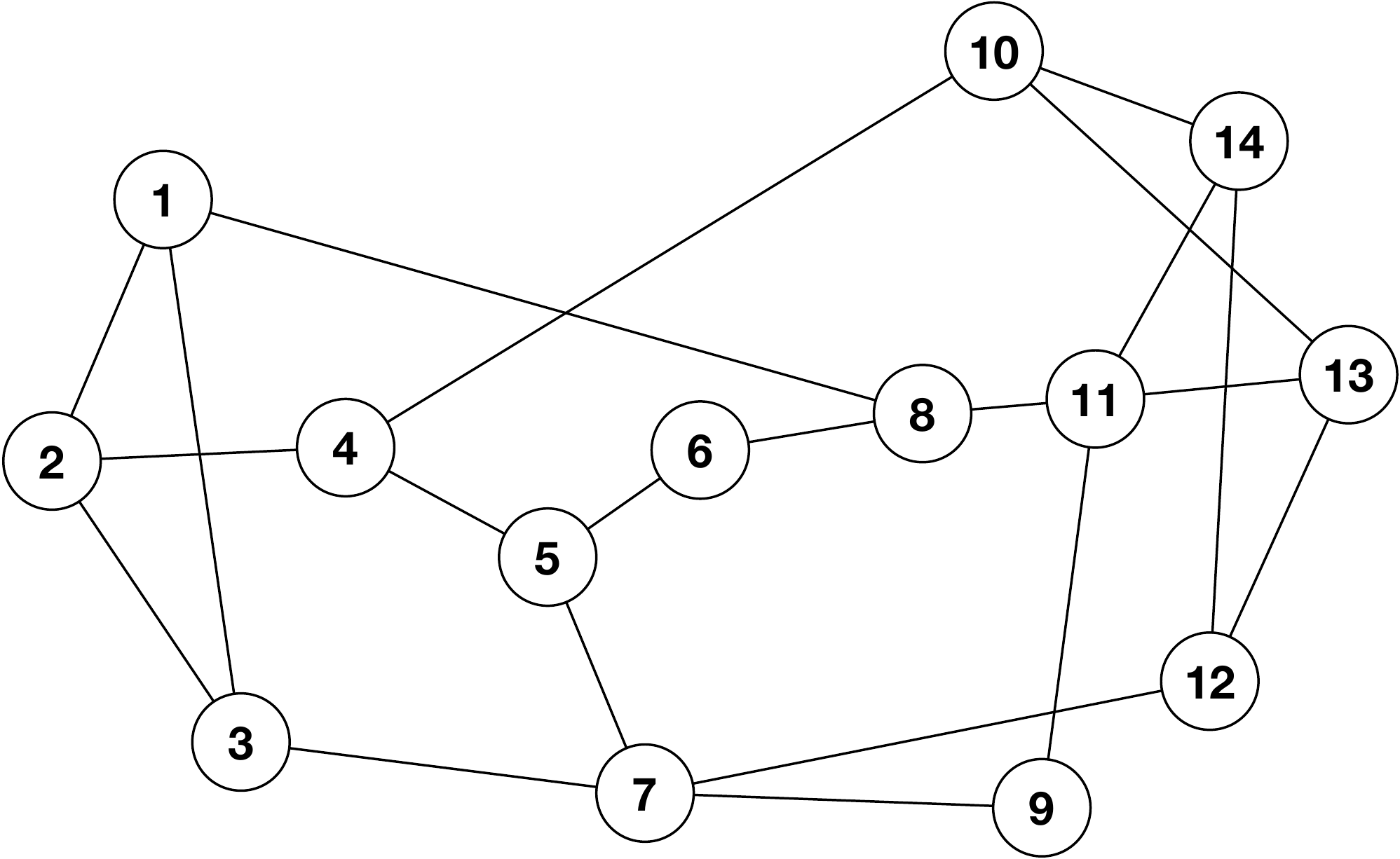}
\caption{NSF}
\label{fig:nsf}
\end{figure}
\begin{figure}[!t]
\includegraphics[scale=0.3]{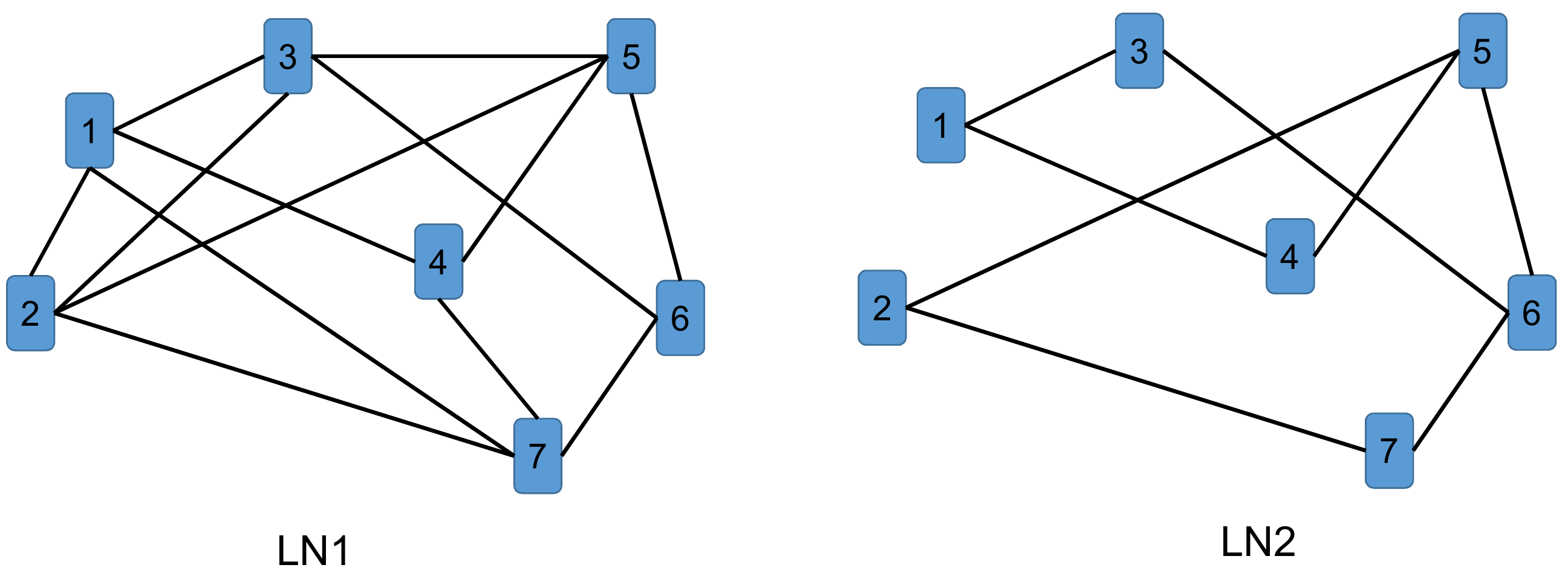}
\caption{LN1 and LN2}
\label{fig:lns}
\end{figure}
We first select NSF network as a small-size physical network 
and create two logical networks denoted as ``LN1'' and ``LN2''. All networks are illustrated in Figs.~\ref{fig:nsf} and~\ref{fig:lns}. Two cross-layer network mappings are created: LN1-over-NSF, and LN2-over-NSF. We apply the survivable cross-layer routing MIP formulation (SUR-TEST) (see Appendix~\ref{app:MIP}) which verifies that LN1-over-NSF is survivable and LN2-over-NSF is non-survivable.
\begin{figure}[!t]
\centering
\includegraphics[scale=0.4]{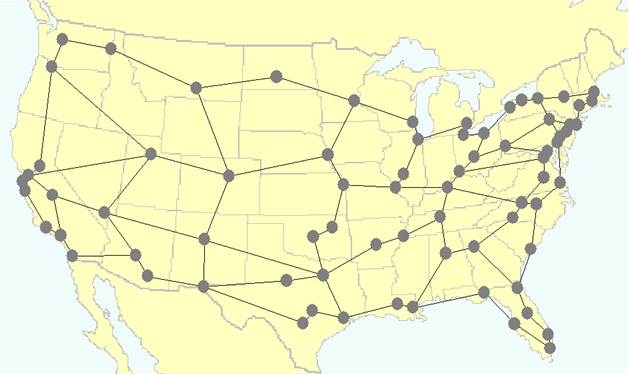}
\caption{CONUS network~\cite{conusNet}}
\label{fig:conus}
\end{figure}
\begin{table}[!h]
\centering
\begin{tabular}{>{\centering\arraybackslash}p{0.9cm}
>{\centering\arraybackslash}p{0.5cm}
>{\centering\arraybackslash}p{0.35cm}
>{\centering\arraybackslash}p{0.35cm}
>{\centering\arraybackslash}p{0.8cm}
>{\centering\arraybackslash}p{0.4cm}
>{\centering\arraybackslash}p{0.3cm}
>{\centering\arraybackslash}p{0.3cm}
>{\centering\arraybackslash}p{0.4cm}
>{\centering\arraybackslash}p{0.4cm}
}\\
\hline\hline
\rule{0pt}{8pt}\multirow{2}{*}{PhyNet} &\multirow{2}{*}{LogNet} &\multirow{2}{*}{Suv} &\multirow{2}{*}{nSuv}&\multirow{2}{*}{FPbRg}&\multirow{2}{*}{uFPb}&\multicolumn{2}{c}{rFPb}&\multicolumn{2}{c}{NumFPb}\\
\rule{0pt}{8pt} & & &&&&Mean&Vrn&uFPb&rFPb\\
\hline
\rule{0pt}{8pt}NSF    &LN1  &1   &0    &[15\%,0\%)&0.1\%&0.5\%&2\%&150&30\\
\rule{0pt}{8pt}NSF    &LN2  &0   &1    &[15\%,0\%)&0.1\%&0.5\%&2\%&150&30\\
\rule{0pt}{8pt}CORONET&CLN1 &9/40&31/40&[15\%,0\%)&-&0.5\%&2\%&-&30\\
\rule{0pt}{8pt}CORONET&CLN2 &7/40&33/40&[15\%,0\%)&-&0.5\%&2\%&-&30\\
\hline\hline
\end{tabular}
\caption{Parameters for testing cases}
\label{tbl:suvNonSuvCase}
\end{table}
\subsubsection{Medium Size Cross-layer Network with CORONET as the Physical Network}
To further validate the scalability of our proposed approach, we select the CORONET network~\cite{saleh2006dynamic} as the physical network, which has 75 nodes, 99 links, and an average nodal degree of 2.6. With CORONET as the physical network, we create 80 logical networks; half of them have nodes randomly selected from 20\% of the physical nodes (denoted as CLN1), and the other half have 30\% (denoted as CLN2).  The average nodal degree for all logical networks is 4. With the logical nodes in CLN1 and CLN2, we generate the cross-layer networks as follows. We first generate a random spanning tree, and then utilize the Erd\H{o}s-R\'{e}nyi random graph model~\cite{erdos1960evolution} to guarantee the connectivity of logical nodes. Finally, random logical-to-physical node mapping are constructed. 
Out of all generated cross-layer networks, we report the number of survivable and unsurvivable cases in Table~\ref{tbl:suvNonSuvCase}, which are all validated by the SUR-TEST MIP formulation.
\subsubsection{Probability of Failure on Physical Links}
The failure probabilities are chosen as follows.
The unified failure probability $\rho$ is selected in the range of $15.0\%\geq \rho > 0\%$ with 0.1\% per step. In total, we have 150 uniform probabilities $[15\%, 14.9\%, \ldots, 0.2\%, 0.1\%]$.

For the random failure probabilities, we generate them based on the normal distribution with the mean from 15.0\% to 0\%, 0.5\% per step, and the variance is 2\%. Note here that the randomly generated probabilities are selected if less than 100\%. In total, we have 30 random failure probabilities.
\subsubsection{Testing Cases}
Parameters to construct all simulation cases are presented in Table~\ref{tbl:suvNonSuvCase}, in which ``PhyNet'', ``LogNet'',``Suv'',``nSuv'',``FPbRg'',``uFPb'' denote the physical network, logical network, the number of survivable and non-survivable cases, the range of failure probabilities, and the incremental step width of unified failure probability. Let ``rFPb''``Mean'',``Vrn'' be random failure probability, mean/step width, and variance;
and let ``NumFPb'',``uFPb'', and ``rFPb'' indicate the total number of unified failure probabilities, and the total number of random failure probabilities for each cross-layer network. The simulation results for all these cases are grouped by the failure probabilities, survivability of the networks, and the size of networks (small and medium).

The performance of the simulations with unified failure probability is only reported with two cross-layer networks, namely LN1-over-NSF and LN2-over-NSF, where the NSF network in both of them is associated with the 150 failure probabilities mentioned above. Similarly, we also evaluate each of them with randomly generated failure probabilities.

Since the unified failure probability is a special case of the random failure probability, we only consider random failure probabilities in the medium-size cross-layer networks based on the generation of CLN1-over-CORONET and CLN2-over-CORONET. 30 failure probabilities are generated for each of the medium-size networks, and these testing cases are grouped and reported by the mean of failure probability and its survivability.
Note here that as part of the validation, LN1-over-NSF and the survivable medium-size networks are expected to reach 100\% survivable probability regardless of their failure probabilities.
\subsection{Simulation Results}\label{subsec:cResults}
In this section, we report the simulation results based on the testing cases described above.
\subsubsection{Small-size Cross-layer Networks}\label{subsubsec:cRsl_small}
The computational results for the survivable probability of the maximal protecting spanning tree and base protecting spanning tree set are denoted as ``MaxPrctTree'' and ``BasePrctTreeSet'', respectively.

\begin{figure}[!t]
\begin{subfigure}[b]{0.25\textwidth}
    \includegraphics[scale=0.26]{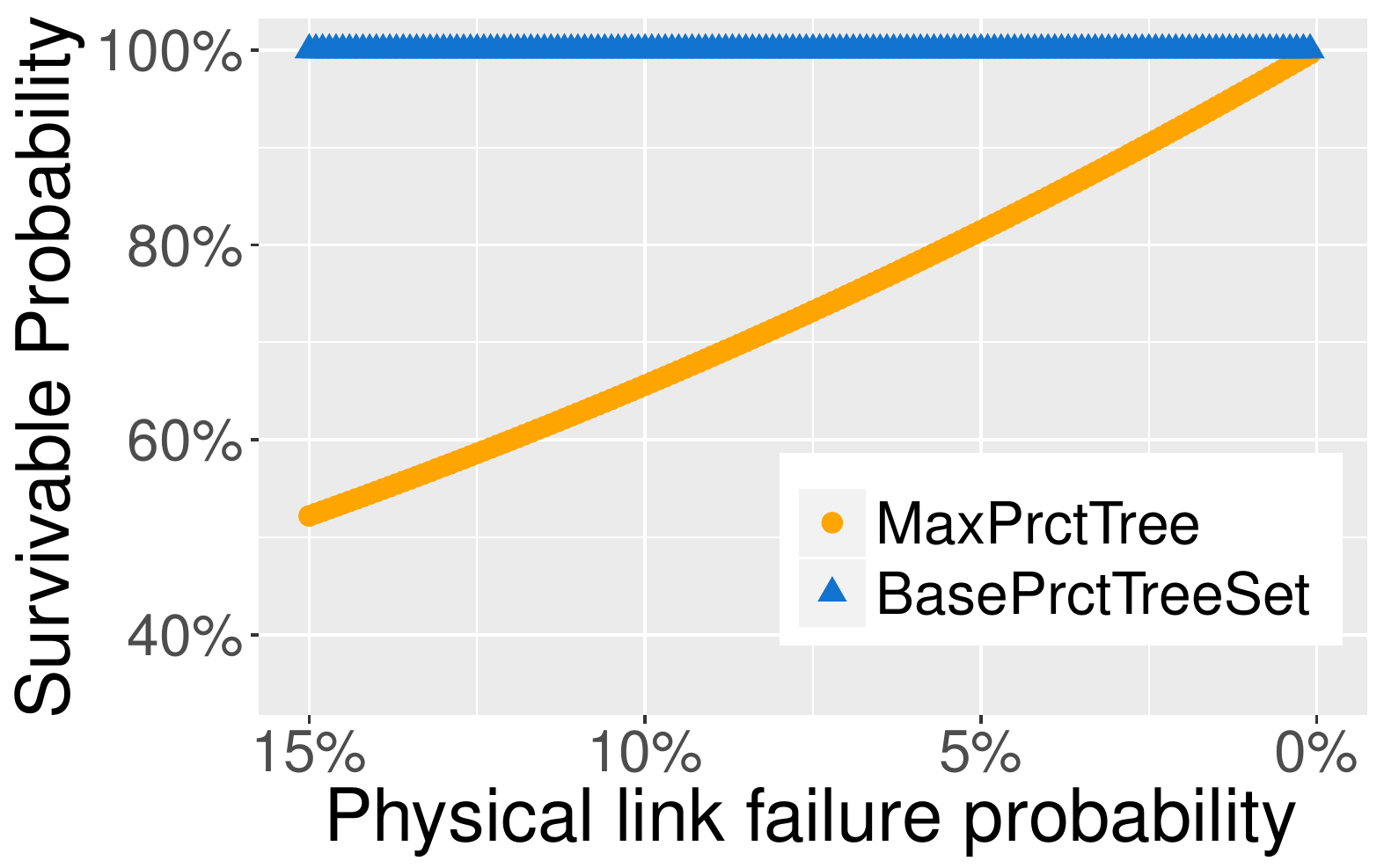}
    \caption{LN1-over-NSF}
    \label{subfig:suvUniNSF}
\end{subfigure}%
\begin{subfigure}[b]{0.25\textwidth}
    \includegraphics[scale=0.26]{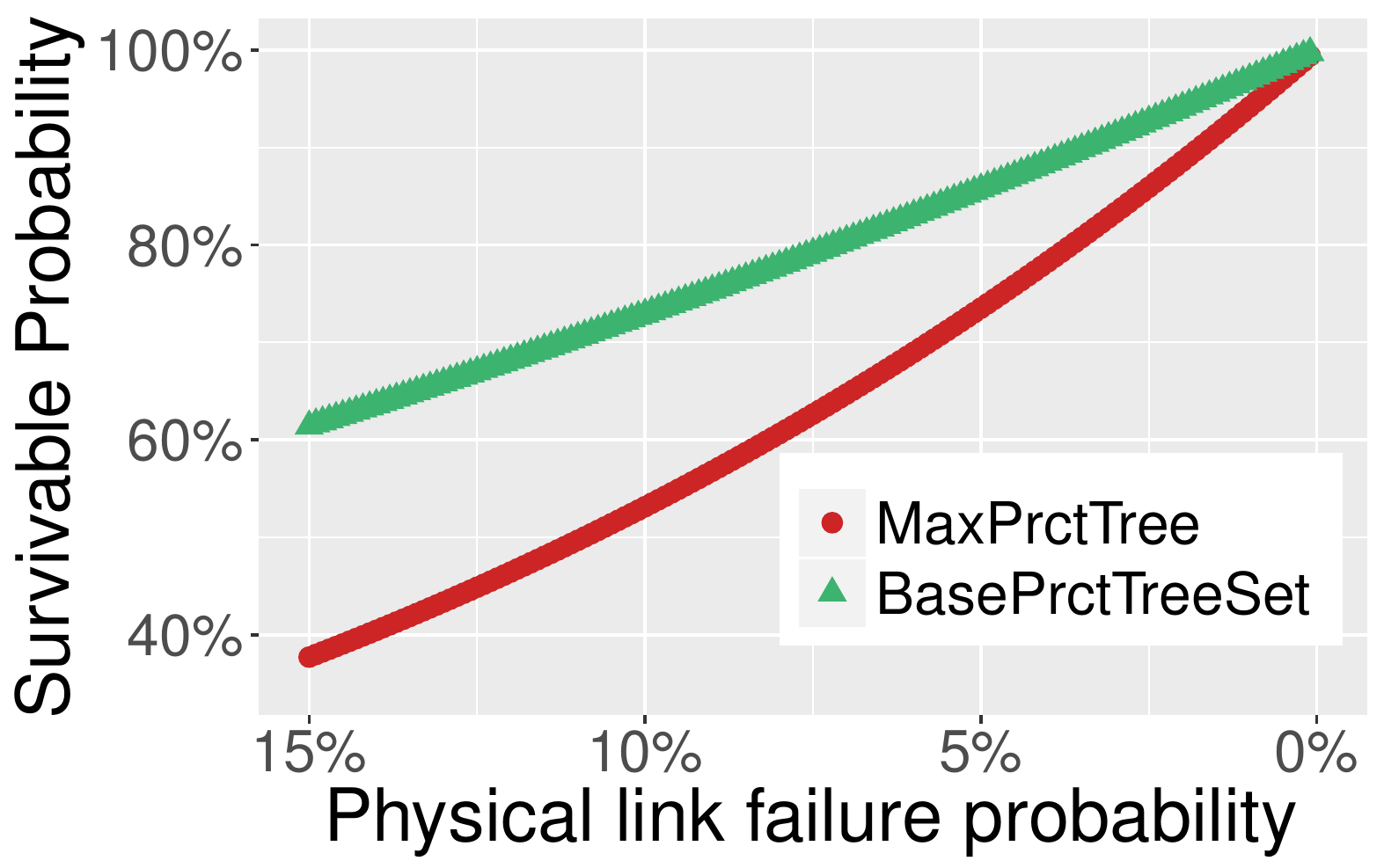}
    \caption{LN2-over-NSF}
    \label{subfig:nonsuvUniNSF}
\end{subfigure}
\caption{Survivable probability with unified failure probability for small-size cross-layer networks}
\label{fig:uniNSF}
\end{figure}

\begin{figure}[!t]
\begin{subfigure}[b]{0.25\textwidth}
    \includegraphics[scale=0.26]{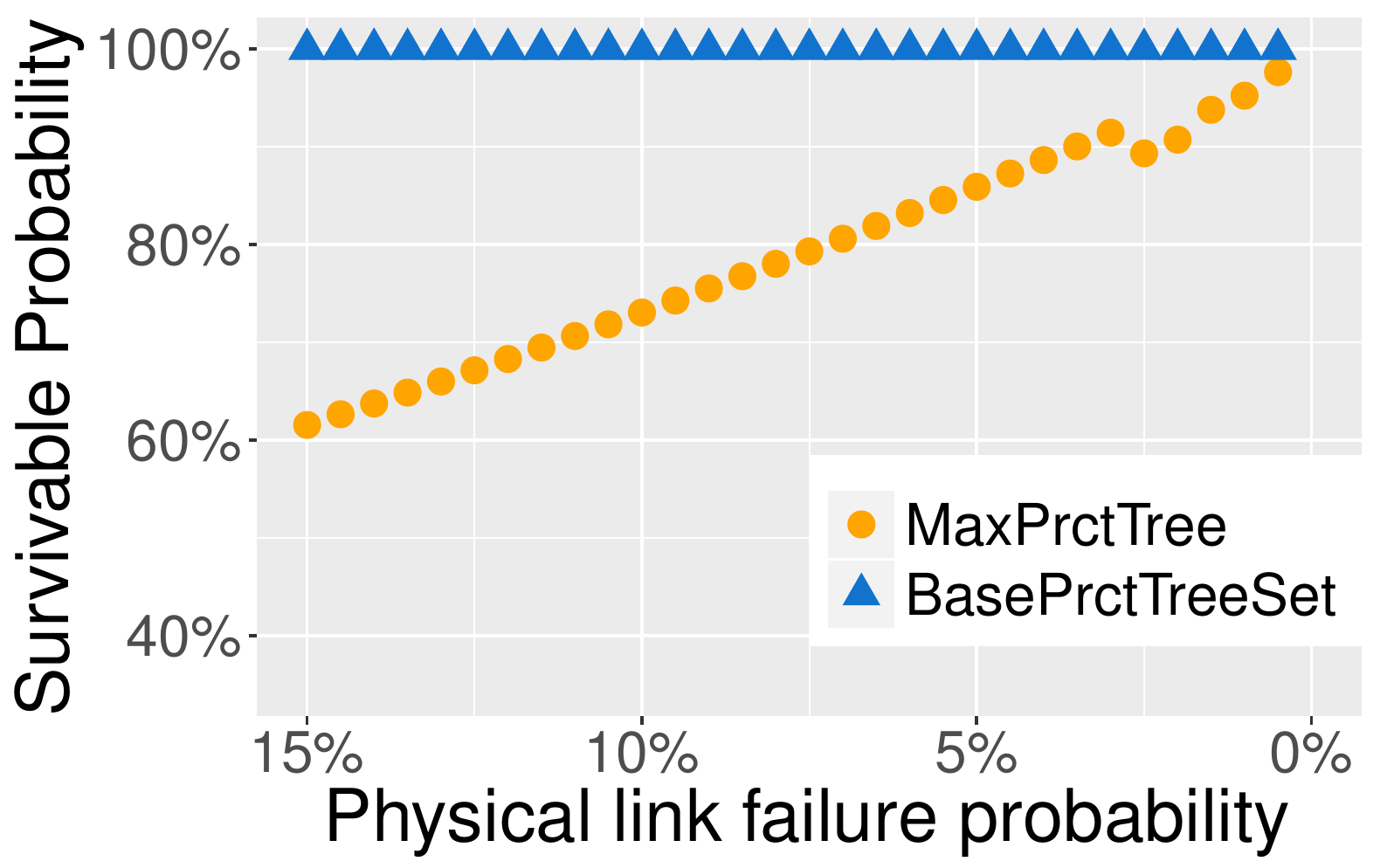}
    \caption{LN1-over-NSF}
    \label{subfig:suvUniNSF}
\end{subfigure}%
\begin{subfigure}[b]{0.25\textwidth}
    \includegraphics[scale=0.26]{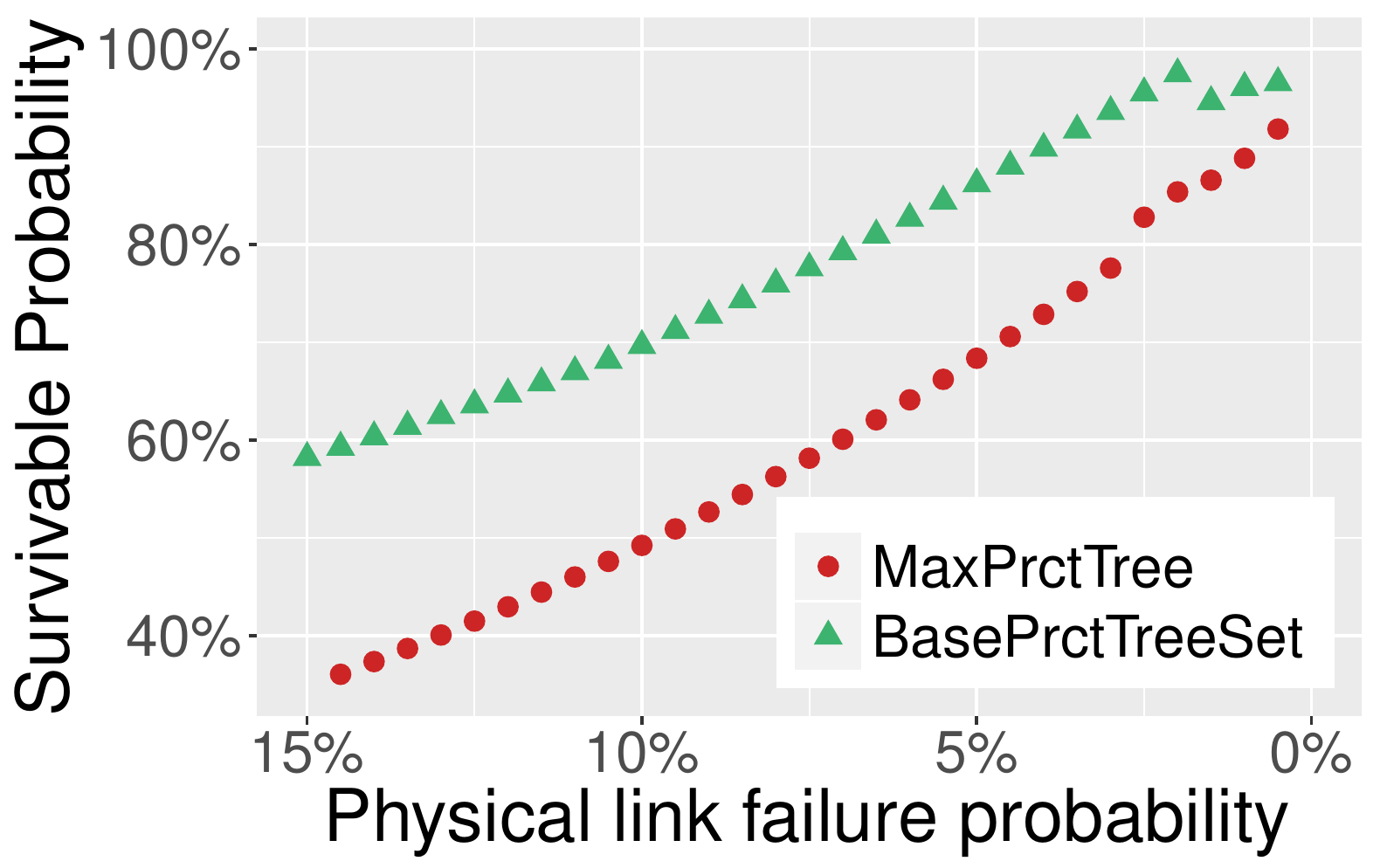}
    \caption{LN2-over-NSF}
    \label{subfig:nonsuvUniNSF}
\end{subfigure}
\caption{Survivable probability with random failure probability for small-size cross-layer networks}
\label{fig:ranNSF}
\end{figure}

Figures~\ref{fig:uniNSF} and~\ref{fig:ranNSF} illustrate the survivable probability of MaxPrctTree and BasePrctTreeSet for LN1-over-NSF and LN2-over-NSF with unified and random failure probabilities, respectively.
These results validate our proposed solution approach as follows: (1) all testing cases for the survivable LN1-over-NSF network are with 100\% survivable probability through the base protecting spanning tree set, regardless of the values/distribution of the failure probabilities; (2) with the unified failure probability, the minimal number of physical links shared by the trees in the base protecting spanning tree set, denoted as $k_{\text{min}}$, is 3 in the LN1-over-NSF network. We validate that the survivable probability obtained by the base protecting spanning tree approach, illustrated in Fig.~\ref{fig:uniNSF}, which matches $(1-\rho)^{k_{\text{min}}}$. These results provide the numerical proof for Theorem~\ref{thm:failProbMinLk}; (3) the curves of survivable probabilities of MaxPrctTree and BasePrctTreeSet over randomly generated failure probabilities are not smooth. But in general, their survivable probabilities are still monotonically increasing while the mean of the failure probability decreases. In other word, as expected, the lower the failure probability, the higher the survivable probability of MaxPrctTree and BasePrctTreeSet are achieved; and (4) the base protecting spanning tree approach works for both survivable and non-survivable cross-layer networks.

\begin{figure}[!t]
\begin{subfigure}[b]{0.25\textwidth}
    \includegraphics[scale=0.26]{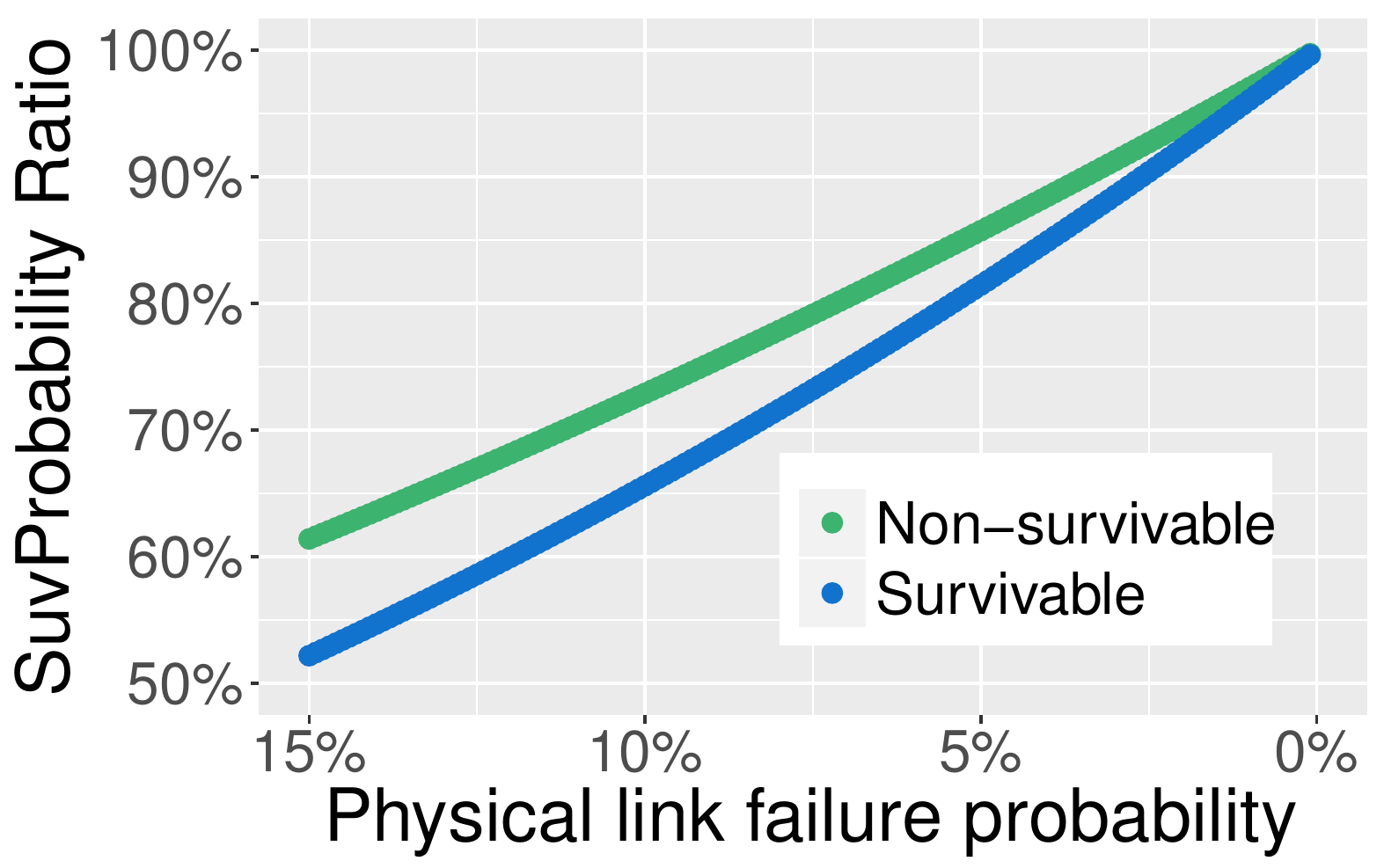}
    \caption{Unified physical link failure probability}
    \label{subfig:suvUniNSF}
\end{subfigure}%
\begin{subfigure}[b]{0.25\textwidth}
    \includegraphics[scale=0.26]{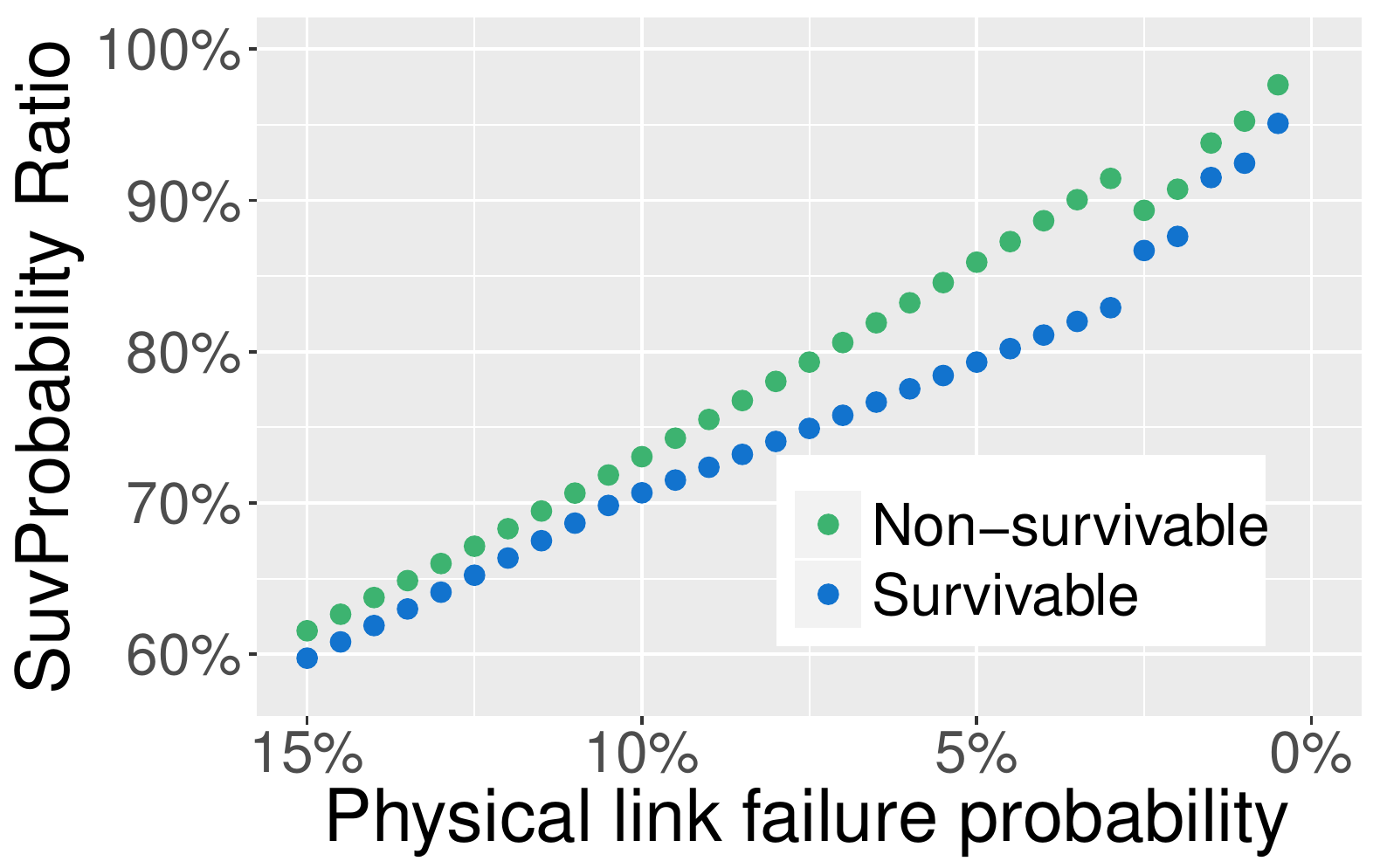}
    \caption{Random physical link failure probability}
    \label{subfig:nonsuvUniNSF}
\end{subfigure}
\caption{Survivable probability ratio between the maximal protecting spanning tree and a base protecting spanning tree set}
\label{fig:NSFratio}
\end{figure}

To demonstrate that the MaxPrctTree may be used to estimate the lower bound of survivable probability, Fig.~\ref{fig:NSFratio} illustrates the ratio of the maximal protecting spanning tree's survivable probability to the survivable probability of a cross-layer network (through a base protecting spanning tree set).
These results show that for all testing cases, the survivable probability of BasePrctTreeSet is higher than that of MaxPrctTree. The lower the probability of failure on physical links, the better the lower bound estimation the maximal protecting spanning tree can provide. With up to 15\% of the average failure probability, the lower bound estimation is higher than $\frac{1}{2}$ of the survivable probability of all the generated cross-layer networks.

\subsubsection{Medium-size Cross-layer Networks}
\begin{figure}[!t]
\begin{subfigure}[b]{0.25\textwidth}
    \includegraphics[scale=0.26]{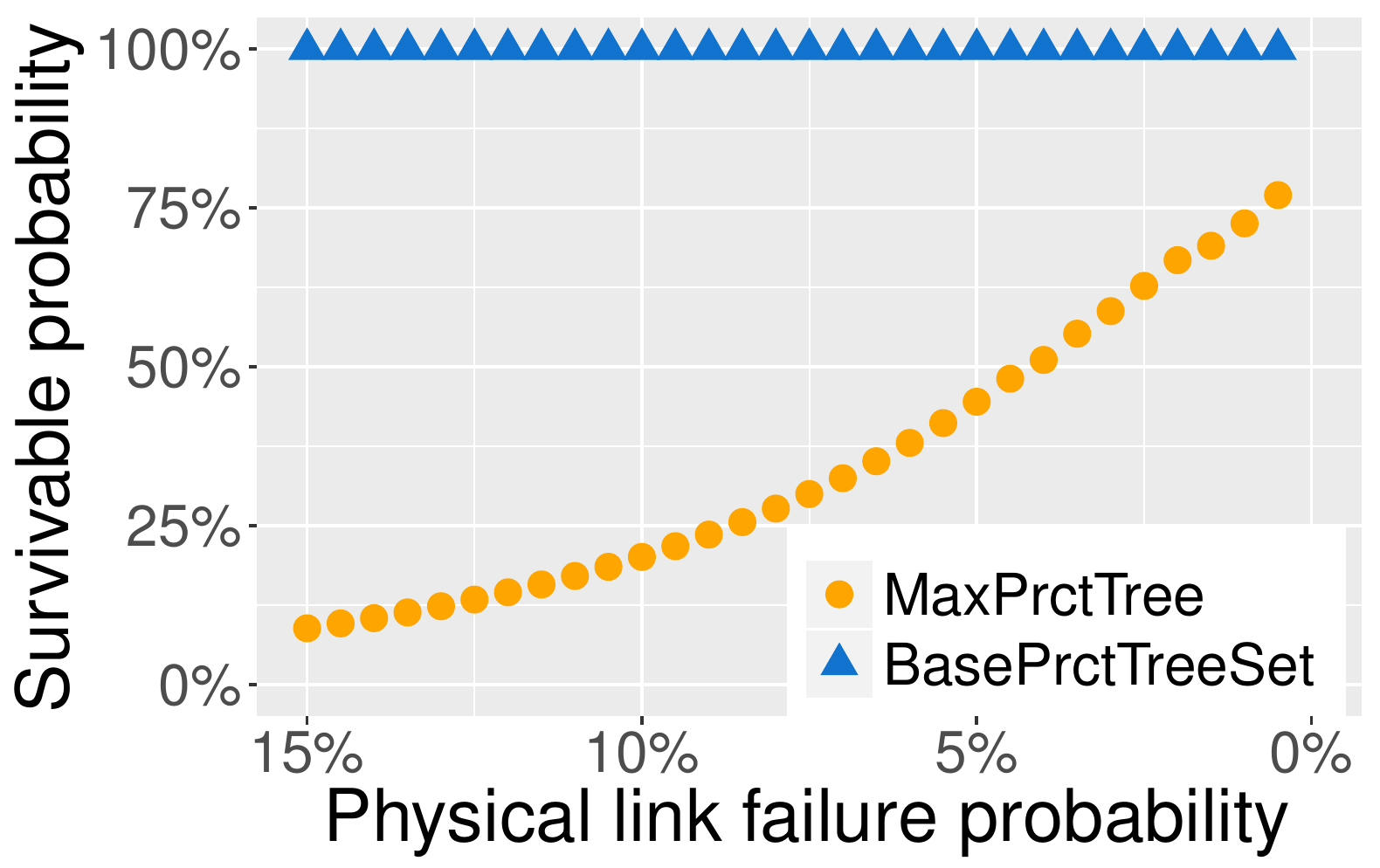}
    \caption{CLN1-over-CONUS}
    \label{subfig:SuvUniConus1}
\end{subfigure}%
\begin{subfigure}[b]{0.25\textwidth}
    \includegraphics[scale=0.26]{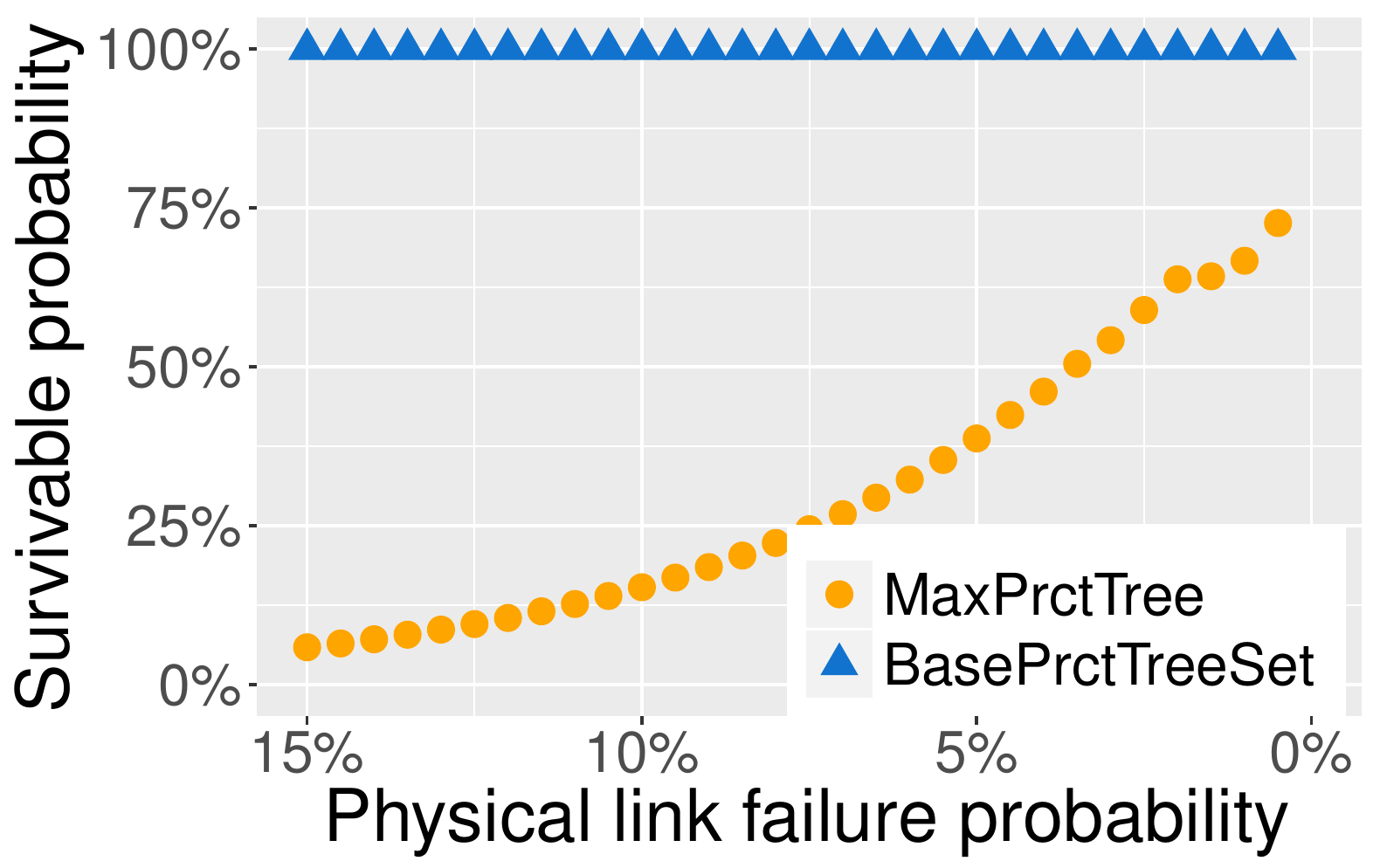}
    \caption{CLN2-over-CONUS}
    \label{subfig:SuvUniConus2}
\end{subfigure}
\caption{Survivable probability with unified failure probability for medium-size survivable cross-layer networks}
\label{fig:ranSuvConus}
\end{figure}

\begin{figure}[!t]
\begin{subfigure}[b]{0.25\textwidth}
    \includegraphics[scale=0.26]{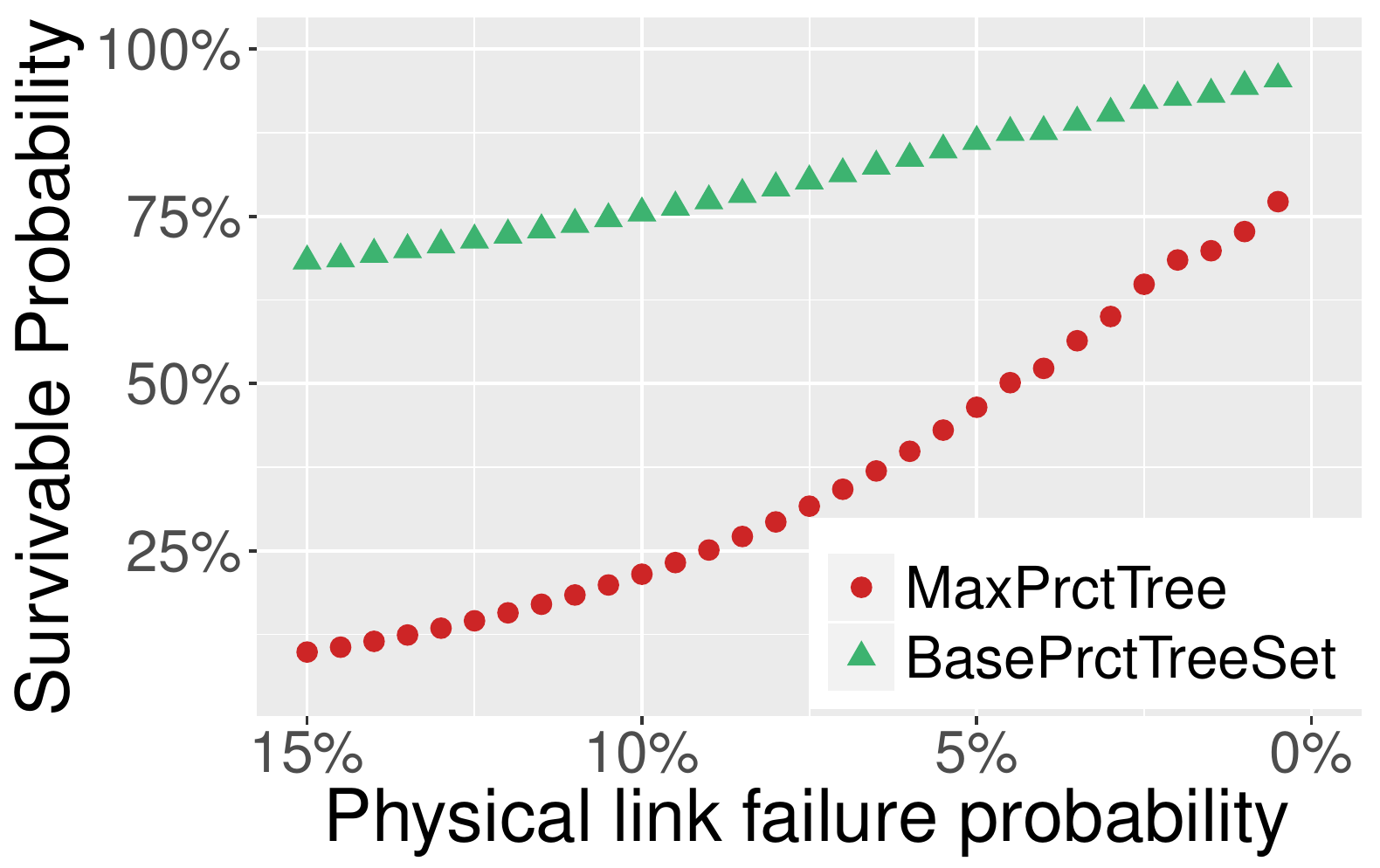}
    \caption{CLN1-over-CONUS}
    \label{subfig:nonSuvRanConus1}
\end{subfigure}%
\begin{subfigure}[b]{0.25\textwidth}
    \includegraphics[scale=0.26]{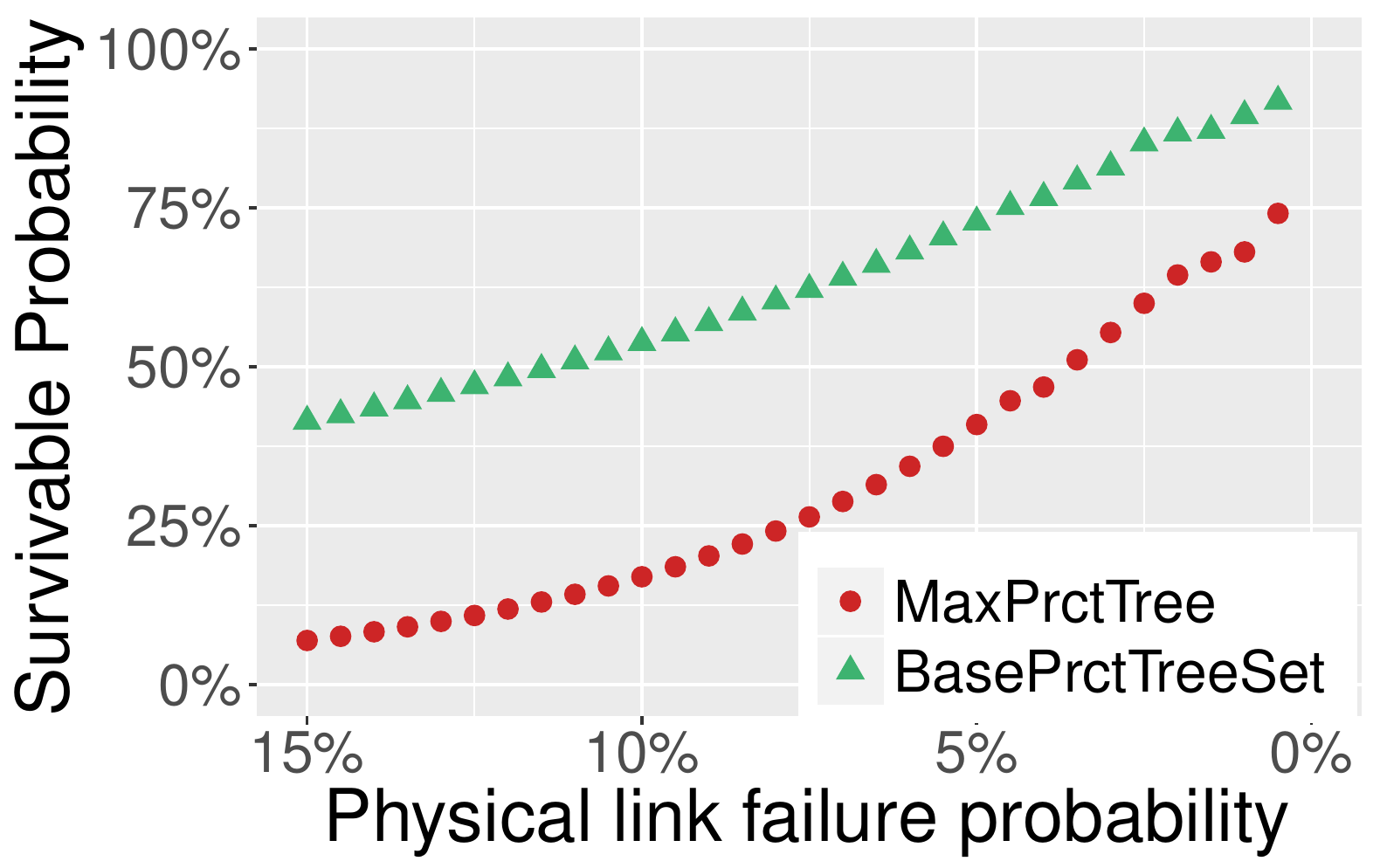}
    \caption{CLN2-over-CONUS}
    \label{subfig:nonSuvRanConus2}
\end{subfigure}
\caption{Survivable probability with random failure probability for medium-size non-survivable cross-layer networks}
\label{fig:nonSuvRanConus}
\end{figure}

\begin{figure}[!t]
\begin{subfigure}[b]{0.25\textwidth}
    \includegraphics[scale=0.26]{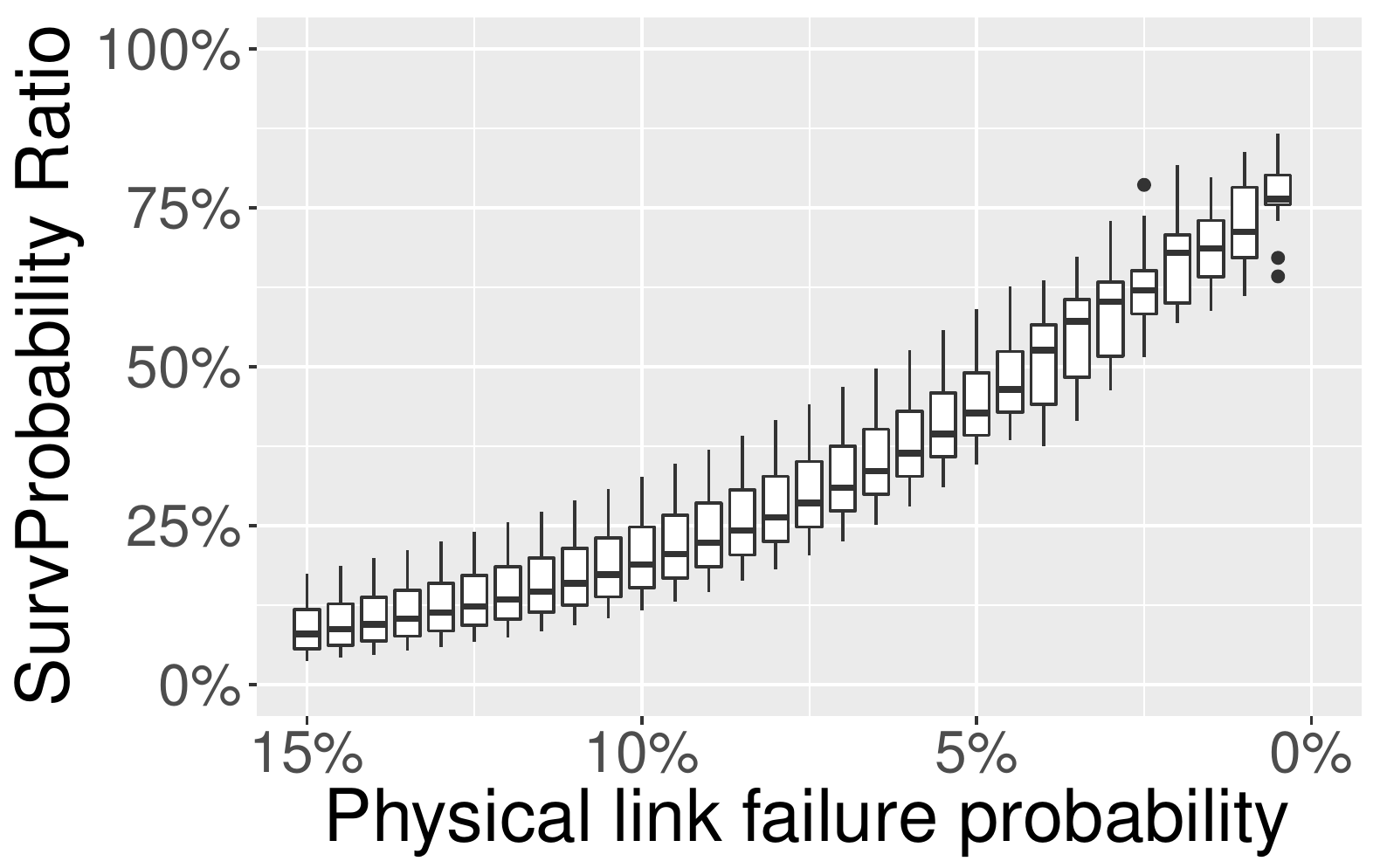}
    \caption{CLN1-over-CONUS}
    \label{subfig:nonSuvRanConus1}
\end{subfigure}%
\begin{subfigure}[b]{0.25\textwidth}
    \includegraphics[scale=0.26]{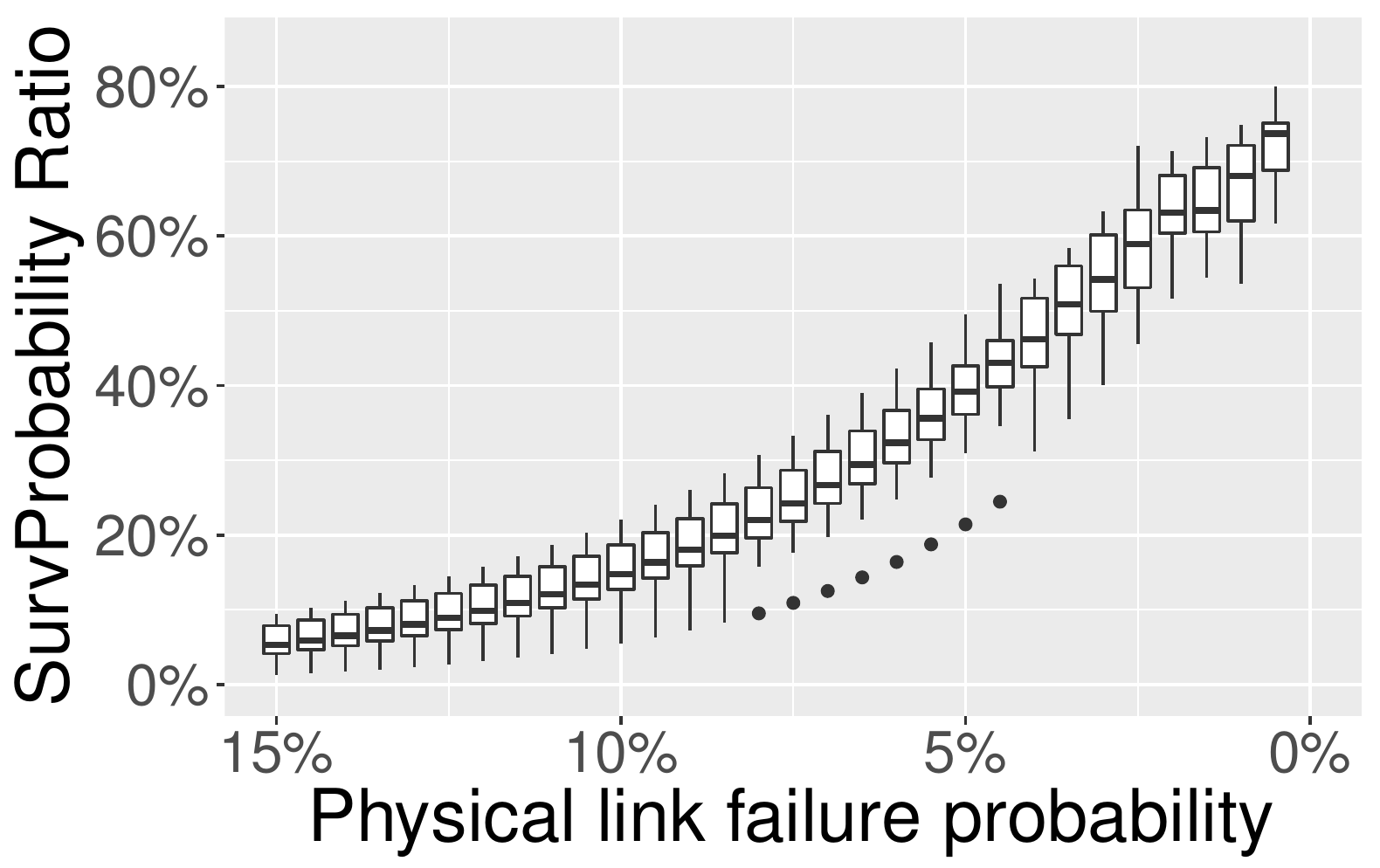}
    \caption{CLN2-over-CONUS}
    \label{subfig:nonSuvRanConus2}
\end{subfigure}
\caption{Survivable probability ratio of MaxPrctTree to BasePrctTreeSet for survivable medium size networks}
\label{fig:SuvRanConusRatio}
\end{figure}

\begin{figure}[!t]
\begin{subfigure}[b]{0.25\textwidth}
    \includegraphics[scale=0.26]{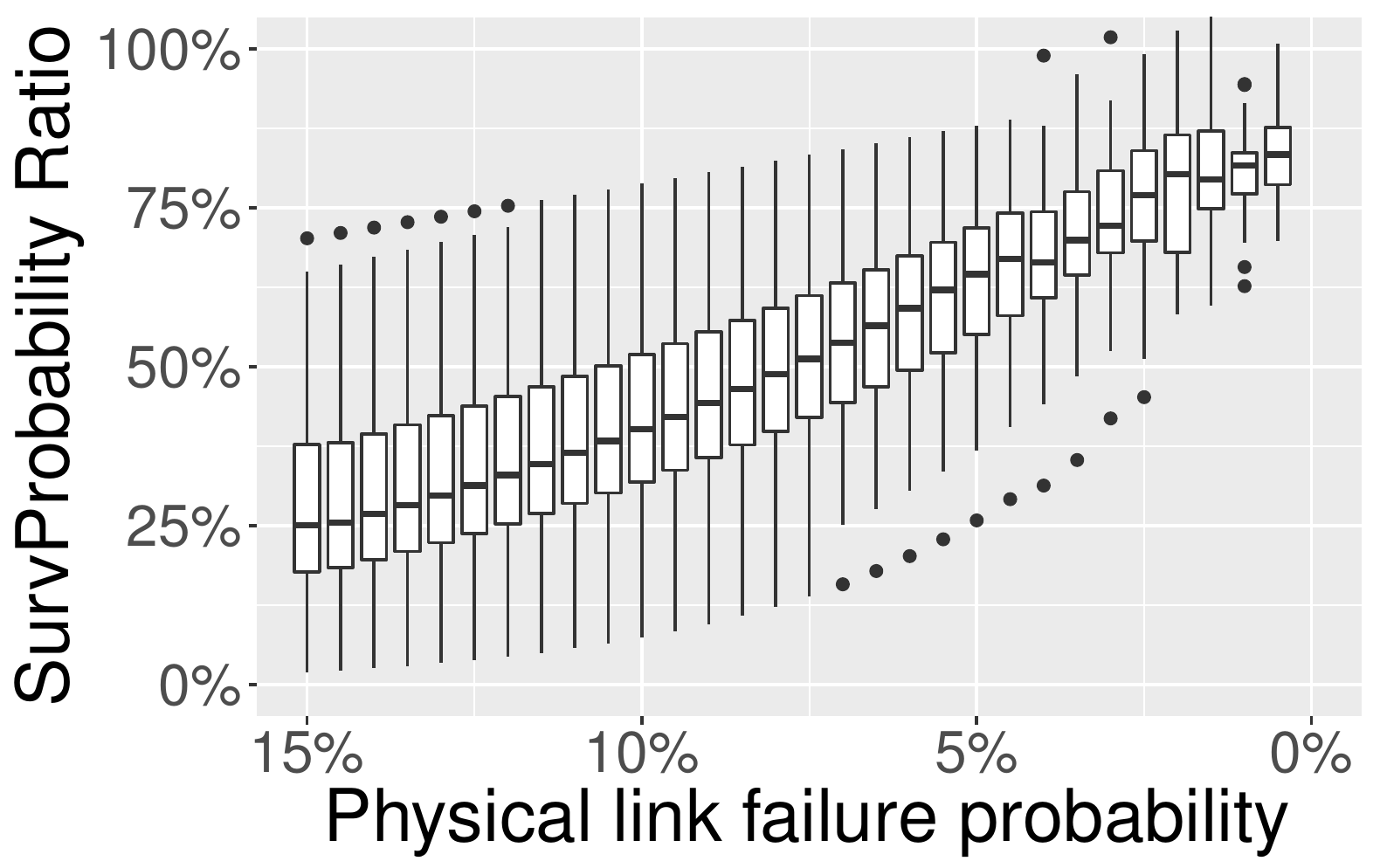}
    \caption{CLN1-over-CONUS}
    \label{subfig:nonSuvRanConus1}
\end{subfigure}%
\begin{subfigure}[b]{0.25\textwidth}
    \includegraphics[scale=0.26]{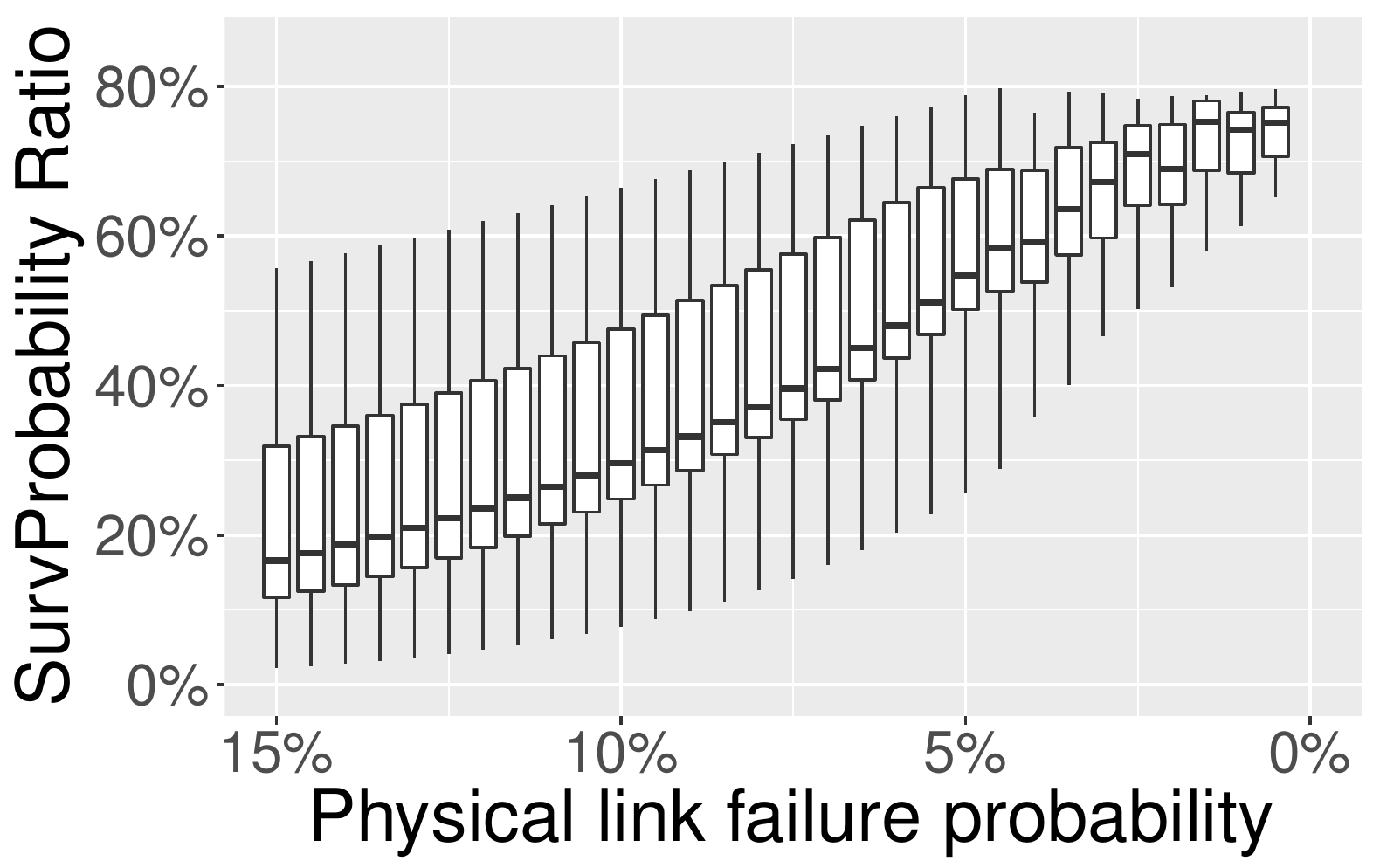}
    \caption{CLN2-over-CONUS}
    \label{subfig:nonSuvRanConus2}
\end{subfigure}
\caption{Survivable probability ratio of MaxPrctTree to BasePrctTreeSet for non-survivable medium-size networks}
\label{fig:nonSuvRanConusRatio}
\end{figure}

Figs.~\ref{fig:ranSuvConus} and~\ref{fig:nonSuvRanConus} illustrate the survivable probability of the survivable and non-survivable cross-layer networks, respectively, where each testing instance is with random failure probabilities on physical links.
Figs.~\ref{fig:SuvRanConusRatio} and~\ref{fig:nonSuvRanConusRatio} present the survivable probability ratio of MaxPrctTree to BasePrctTreeSet for all network instances in box plots, which are grouped by their respective failure probabilities.
These results further validate our proposed solution approaches that (1) for all survivable cases (verified by the SUR-TEST formulation), our approaches produce 100\% survivable probability;
(2) the survivable probability of BasePrctTreeSet is higher than that of MaxPrctTree for all testing cases; (3) with larger logical networks (CLN2), more physical links are utilized by logical link mappings, which bring down the survivable probability of MaxPrctTree significantly compared with the smaller-size ones.

The computational time of all MIP formulations are finished within 15 minutes, thus our proposed solution approaches can produce results effectively at least for the medium-size networks.

We also observe some interesting facts which may direct our future studies on network properties. (1) The average survivable probability ratio for both survivable and non-survivable networks is monotonically increasing when failure probability decreases. (2) When failure probability decreases, gaps of the survivable probability ratios for all tested survivable networks are increasing (see Fig.~\ref{fig:SuvRanConusRatio}; and the gaps of survivable probability ratios for all tested non-survivable networks are decreasing (see  Fig.~\ref{fig:nonSuvRanConusRatio}). (3) In general, the computational time for the survivable cases is higher than that of the non-survivable ones.